\documentclass[aps,prd,twocolumn,superscriptaddress,nofootinbib]{revtex4-1}

\usepackage{latexsym}
\usepackage{amsmath}
\usepackage{amssymb}
\usepackage{amsfonts}
\usepackage{bm}
\usepackage{physics}

\usepackage{color}
\definecolor{purple}{rgb}{0.5,0,0.5}
\definecolor{blue}{rgb}{0.0,0,0.9}
\definecolor{prdblue}{rgb}{0.133,0.118,0.498}
\usepackage[colorlinks=true, pdfstartview=FitV, linkcolor=prdblue, citecolor= prdblue, urlcolor=prdblue]{hyperref}

\usepackage{supertabular} 
\usepackage{placeins}
\usepackage{epsfig}
\usepackage{graphicx}

\def\tstrut{\vrule height3.25ex depth0pt width0pt} 

\begin{document}


\title{An algebraic model to study the internal structure of pseudo-scalar mesons with heavy-light quark content}


\author{B. Almeida-Zamora}
\email[]{bilgai\_almeidaz@hotmail.com}
\affiliation{Departamento de Investigaci\'on en F\'isica, Universidad de Sonora, Boulevard Luis Encinas J. y Rosales, 83000, Hermosillo, Sonora, Mexico.}
\affiliation{Departamento de Sistemas F\'isicos, Qu\'imicos y Naturales, Universidad Pablo de Olavide, E-41013 Sevilla, Spain}

\author{J.J. Cobos-Mart\'inez}
\email[]{jesus.cobos@unison.mx}
\affiliation{Departamento de F\'isica, Universidad de Sonora, Boulevard Luis Encinas J. y Rosales, 83000, Hermosillo, Sonora, Mexico.}

\author{A. Bashir}
\email[]{adnan.bashir@umich.mx;abashir@jlab.org}
\affiliation{Instituto de F\'sica y Matem\'aticas, Universidad Michoacana de San Nicol\'as de Hidalgo, Morelia, Michoac\'an 58040, M\'exico.}
\affiliation{Theory Center, Jefferson Lab, Newport News, VA 23606, USA}

\author{K. Raya}
\email[]{khepani.raya@dci.uhu.es}
\affiliation{Departmento de Ciencias Integradas, Universidad de Huelva, E-21071 Huelva, Spain.}

\author{J. Rodr\'iguez-Quintero}
\email[]{jose.rodriguez@dfaie.uhu.es}
\affiliation{Departmento de Ciencias Integradas, Universidad de Huelva, E-21071 Huelva, Spain.}

\author{J. Segovia}
\email[]{jsegovia@upo.es}
\affiliation{Departamento de Sistemas F\'isicos, Qu\'imicos y Naturales, Universidad Pablo de Olavide, E-41013 Sevilla, Spain}


\date{\today}

\begin{abstract}
The internal structure of all lowest-lying pseudo-scalar mesons with heavy-light quark content is studied in detail using an algebraic model that has been applied recently, and successfully, to the same physical observables of pseudo-scalar and vector mesons with hidden-flavor quark content, from light to heavy quark sectors.
The algebraic model consists on constructing simple and evidence-based \emph{ans\"atze} of the meson's Bethe-Salpeter amplitude (BSA) and quark's propagator in such a way that the Bethe-Salpeter wave function (BSWF) can then be readily computed algebraically. Its subsequent projection onto the light front yields the light front wave function (LFWF) whose form allows us a simple access to the valence-quark Parton Distribution Amplitude (PDA) by integrating over the transverse momentum squared.
We exploit our current knowledge of the PDAs of lowest-lying pseudo-scalar heavy-light mesons to compute their Generalized Parton Distributions (GPDs) through the overlap representation of LFWFs. From these three dimensional knowledge, different limits/projections lead us to deduce the related Parton Distribution functions (PDFs), Electromagnetic Form Factors (EFFs), and Impact parameter space GPDs (IPS-GPDs).
When possible, we make explicit comparisons with available experimental results and earlier theoretical predictions.
\end{abstract}


\maketitle


\section{Introduction}
\label{sec:intro}

In working towards an understanding of hadrons, we have discovered that they are complicated bound-states of quarks and gluons whose interactions are described by a Poincar\'e invariant quantum non-Abelian gauge field theory; namely, Quantum Chromodynamics (QCD). Whilst the principles and practice of perturbation theory in QCD are fairly well understood, non-perturbative tools such as lattice-regularized QCD and continuum Dyson-Schwinger equations are essential because the gluon and quark degrees-of-freedom used to express the QCD's Lagrangian appear not to be the objects that determine the low-energy ground of QCD where hadrons live. Future experimental facilities worldwide will deliver data that promises to reveal the innermost workings of hadrons in terms of QCD's (effective) elementary excitations: (dressed pseudo-) quarks and gluons. In order to fulfill that promise, phenomenology and theory must be developed accordingly.

Within a continuum Dyson-Schwinger equations approach to the valence-quark and -antiquark bound-state problem in QCD, the Bethe-Salpeter wave function (BSWF) is the key object to compute in order to determine most of the physical observables related to mesons. The BSWF is computed by combining the Dyson-Schwinger equation (DSE) for the quark propagator and the Bethe-Salpeter equation (BSE) for mesons~\cite{Roberts:1994dr, Maris:1997tm, Chang:2009zb, Roberts:2011wy, Binosi:2016rxz, Qin:2020jig, Xu:2022kng}. This formalism has produced a plethora of theoretically interesting and experimentally testable quantities such as mesons masses~\cite{Roberts:2011cf, Roberts:2011wy, Chen:2012qr, Qin:2019oar, Qin:2020jig, Xu:2022kng}, their other static properties~\cite{Yin:2019bxe, Yin:2021uom}, form factors (FFs)~\cite{Chang:2013nia, Raya:2015gva} and parton distribution amplitudes (PDAs)~\cite{Gao:2014bca, Ding:2015rkn, Binosi:2018rht, Serna:2020txe}. However, the calculation of, for example, parton distribution functions (PDFs)~\cite{Ding:2019lwe, Cui:2020tdf}, generalized parton distributions (GPDs)~\cite{Mezrag:2014jka, Raya:2021zrz, Zhang:2021mtn} and transverse momentum distributions (TMDs)~\cite{Shi:2018zqd, Shi:2020pqe, Shi:2022erw, Kou:2023ady} remains a highly non-trivial task. 

Our current understanding of the intricate interplay between the DSE of the quark propagator and the meson's BSE allows us to build a simplified model~\cite{Albino:2022gzs} that enables algebraic manipulations whereas produces reliable predictions of physical observables whose extraction from first principles remains still troublesome~\cite{Albino:2022gzs, Almeida-Zamora:2023rwg}. 

The algebraic model consists on constructing the meson's BSWF in terms of the spectral density function (SDF) that appears in a Nakanishi integral representation of the covariant quark-antiquark bound-state amplitude. Then, the BSWF leads to the derivation of the leading-twist LFWF by merely appealing to the definition of its Mellin moments. The resulting LFWF permits an algebraic connection with the PDA so that the need to specify a SDF is completely circumvented with prior knowledge of the PDA. Moreover, we can extract the GPD in the so-called Dokshitzer-Gribov-Lipatov-Altarelli-Parisi (DGLAP) kinematic region through the overlap representation of LFWFs. Finally, different limits/projections of the GPDs provide related PDFs, EFFs and IPS-GPDs.

We exploit our current knowledge of the PDAs of lowest-lying pseudo-scalar heavy-light mesons~\cite{Binosi:2018rht, Serna:2020txe} to compute all corresponding structure distributions mentioned above using the algebraic model that has been successfully applied to the same physical observables of pseudo-scalar~\cite{Albino:2022gzs} and vector~\cite{Almeida-Zamora:2023rwg} mesons with hidden-flavor quark content, from light to heavy quark sectors.

This manuscript is organized as follows. Section~\ref{sec:AlgMod} describes briefly the algebraic model for the quark propagator and the BSA of heavy-light pseudo-scalar mesons. The same section contains, on one hand, a derivation of the algebraic relation between LFWFs and PDAs and, on the other hand, the associated phenomenology of the computed LFWFs for the $D$, $D_s$, $B$, $B_s$ and $B_c$ mesons. With the LFWFs of lowest-lying heavy-light pseudo-scalar mesons at hand, we proceed to discuss the derivation of the corresponding GPDs in the DGLAP kinematic region in Sec.~\ref{sec:IntStrMes}; moreover, we provide related PDFs, EFFs and IPS-GPDs from different limits/projections of the GPDs. Finally, we provide a brief summary and some concluding remarks in Sec.~\ref{sec:summary}.


\section{Algebraic Model}
\label{sec:AlgMod} 

We briefly describe the algebraic model that was reported first in Ref.~\cite{Albino:2022gzs} and successfully applied to the computation of LFWFs of pseudoscalar~\cite{Albino:2022gzs} and vector~\cite{Almeida-Zamora:2023rwg} mesons with hidden-flavor quark content, from light to heavy quark sectors. This presentation is indeed focused on the case of heavy-light pseudo-scalar mesons.

\subsection{Bethe-Salpeter Wave Function}
\label{subsec:BSWF}

The BSWF encapsulates the internal dynamics of a meson in quantum field theory. For a heavy-light pseudoscalar meson, it can be written as
\begin{equation}
\chi_{0^-}(p_-,P) = S_q(p) \, \Gamma_{0^-}(p_-,P) \, S_{\bar Q}(p-P) \,,
\label{eq:BSWF}
\end{equation}
where $p_{-}=p-P/2$, $P^2=-m_{0^-}^2$ is the negative mass squared of the pseudoscalar meson in euclidean-metric space, $S_{q(\bar Q)}$ is the light-quark (heavy-antiquark) propagator and $\Gamma_{0^-}$ is the BSA, where $J^P=0^-$ indicates explicitly the meson's spin-parity quantum numbers.

Within our algebraic model, the expressions for the quark (antiquark) propagator and BSA are given by
\begin{align}
S_{q(\bar Q)}(p) &= \Big(-i\gamma\cdot p + M_{q(\bar Q)} \Big) \Delta(p^2, M_{q(\bar Q)}^2) \,, \label{eq:Sq} \\
N_{0^-} \Gamma_{0^-}(p,P) &= i \gamma_{5} \int_{-1}^{1} dw \; \rho_{0^-}(w) \Big[\hat{\Delta}(p_w^2,\Lambda_w^2) \Big]^\nu \label{eq:BSA} \,,
\end{align}
where $\Delta(s,t)\equiv (s+t)^{-1}$ and $\hat{\Delta}(s,t)\equiv t\Delta(s,t)$. Besides, $N_{0^-}$ is a normalization constant, $M_{q(\bar Q)}$ is the dynamically-dressed quark (antiquark) mass, $\rho_{0^-}(w)$ denotes the SDF which defines the point-wise behavior of the BSA, $p_\omega=p+\frac{\omega}{2}P$ and 
\begin{align}
\Lambda_w^2 \equiv \Lambda^2(w) &= M_q^2 + \frac{1}{2} (M_{\bar Q}^2 - M_{q}^2) (1-w) \nonumber \\
&
- \frac{1}{4} m_{0^-}^2 (1 - w^2) \,.
\label{eq:Lambda2w}
\end{align}
As already noticed in Ref.~\cite{Albino:2022gzs}, the $\omega$-dependence of $\Lambda$ leads to a simplification in relevant integrals, providing closed algebraic expressions that relate different structure distributions. Moreover, Eq.~\eqref{eq:Lambda2w} has some additional striking features that deserve to be highlighted. First, a constant term $M_q^2$ is retained, it is inherited from kindred models~\cite{Raya:2022eqa, Raya:2021zrz, Zhang:2021mtn, Chavez:2021llq, Chavez:2021koz, Chouika:2017dhe, Chouika:2017rzs, Mezrag:2016hnp, Mezrag:2014jka, Xu:2018eii} that have been employed successfully to compute an array of GPD-related distributions. Second, the linear term in $w$ is added in order to study mesons with different flavored quark and antiquark content. And, third, all coefficients in Eq.~\eqref{eq:Lambda2w} are chosen in such a way that the positivity of $\Lambda_\omega^2$ is guaranteed; in particular, one must find sensible values for the constituent masses to fulfill the inequality $|M_{\bar{Q}}-M_q|<m_{0^-}<M_{\bar Q}+M_q$.

It is worth noting herein that the parameter $\nu>-1$ controls the asymptotic behavior of the BSA which must be ultraviolet finite since it resembles the wave function of a bound state~\cite{Maris:1999nt}; therefore, $\nu$ does not control any possible divergence but fits the asymptotic trend of the meson's BSWF. The value $\nu=1$ is used herein since it has been demonstrated that it yields the correct power law of the asymptotic behavior for mesons~\cite{Roberts:1994dr} and, in particular, $\nu=1$ recovers the results in Refs.~\cite{Chavez:2021koz, Chavez:2021llq, Chouika:2017rzs, Mezrag:2016hnp, Mezrag:2014jka}.

Combining Eqs.~\eqref{eq:BSWF} to~\eqref{eq:BSA}, a Nakanishi integral representation (NIR) of the BSWF can be obtained:
\begin{align}
N_{0^-} \chi_{0^-}(p_-,P) &= {\cal M}_{q,\bar Q}(p,P) \nonumber \\
&
\times \int_{-1}^{1} dw \; {\cal D}_{q,\bar Q}^\nu(p,P) \tilde{\rho}_{0^-}^\nu(w) \,,
\label{eq:NIR}
\end{align}
where the function ${\cal M}_{q,\bar Q}(p,P)$ is given by
\begin{align}
{\cal M}_{q,\bar Q}(p,P) &= (-i\gamma\cdot p + M_q) \, i\gamma_{5} \, (-i\gamma\cdot (p-P) + M_{\bar Q}) \nonumber \\
&
= - \gamma_5 \Big\{ M_q \gamma\cdot P + (M_{\bar Q}-M_q) \gamma\cdot p \nonumber \\
&
+ \sigma_{\mu\nu} p_\mu P_\nu - i \big[p\cdot (p-P) + M_qM_{\bar Q} \big] \Big\} \,,
\end{align}
the profile distribution $\tilde{\rho}_{0^-}^\nu(w)$ is defined in terms of the SDF as
\begin{equation}
\tilde{\rho}_{0^-}^\nu(w) = \Lambda_w^{2\nu} \, \rho_{0^-}(w) \,,
\end{equation}
and the product of three denominators,
\begin{align}
\mathcal{D}_{q,\bar Q}^\nu(p,P) &= \Delta\left(p^2,M_q^2\right) \Delta\left(p^2_{w},\Lambda_\omega^2\right)^{\nu} \nonumber \\
&
\times \Delta\left((p-P)^2,M_{\bar Q}^2\right) \,,
\end{align}
can be combined using standard Feynman parametrization techniques to arrive at
\begin{align}
{\cal D}_{q,\bar Q}^\nu(p,P) &= \nu (\nu+1) \int_0^1 d\beta \int_{\frac{1}{2}(\beta-1)(w-1)}^{\frac{1}{2}\left[(w+1)\beta-(w-1)\right]} d\alpha \nonumber \\
&
\times \frac{(1-\beta)^{\nu-1}}{\left[(p-\alpha P)^2 + \Lambda_{1-2\alpha}^2\right]^{\nu+2}} \,.
\label{eq:propD}
\end{align}

Finally, suitable changes of variables and subsequent rearrangements in the order of integrations yield the following expression for the BSWF:
\begin{align}\label{BSWF}
N_{0^-} \chi(p_-,P) &= \mathcal{M}_{q,\bar{Q}}(p,P) \int_{0}^{1} d\alpha \, \mathcal{F}_{0^-}^\nu(\alpha,p,P) \,,
\end{align}
where
\begin{align}
&
\mathcal{F}_{0^-}^\nu(\alpha,p,P) = 2^\nu (\nu+1) \nonumber \\
&
\times \left[ \int_{-1}^{1-2\alpha}dw \left( \frac{\alpha}{1- w}\right)^\nu + \int_{1-2\alpha}^1dw \left(\frac{1- \alpha}{1-w}\right)^\nu \right] \nonumber \\
&
\times \frac{\tilde{\rho}_{0^-}^\nu (w)}{\left[(p-\alpha P)^2 + \Lambda_{1-2\alpha}^2\right]^{\nu+2}} \,.  
\end{align}

\subsection{Light-front Wave Function}
\label{subsec:LFWF}

The LFWF of a pseudoscalar meson can be extracted from the following light-front projection of the BSWF:
\begin{align}
\psi_{0^-}(x,p_\perp^2) &= \Tr \int \frac{d^2 p_\parallel}{\pi} \delta (n\cdot p - x n\cdot P) \nonumber \\
&
\times \gamma_5 (\gamma \cdot n) \chi_{0^-} (p_-, P) \,,
\end{align}
where the trace is taken over color and spinor indices, $n$ is a light-like four-vector, such that $n^2 = 0$ and $n\cdot P = - m_{0^{-}}$, the variable $x$ corresponds to the light-front momentum fraction carried by the quark and the 4-momentum integral has been decomposed as
\begin{equation}
\int \frac{d^4p}{(2\pi)^4} = \int \frac{d^2p_\perp}{16\pi^3} \int \frac{d^2p_\parallel}{\pi} \,.
\end{equation}

We define the Mellin's moments of the pseudoscalar meson's LFWF as
\begin{equation}
\expval{x^m}_{\psi_{0^-}} = \int_0^1 dx \; x^m \, \psi_{0^-} (x, p_\perp^2) \,,
\end{equation}
in such a way that, performing the integral over $x$, the Mellin's moments of the distribution become
\begin{align}
\expval{x^m}_{\psi_{0^-}} &= \Tr \int \frac{d^2 p_\parallel}{\pi} \frac{1}{n \cdot P} \left( \frac{n \cdot p}{n \cdot P}\right)^m \nonumber \\
&
\times \gamma_5 (\gamma \cdot n) \chi_{0^-} (p_{-}, P) \,.
\label{eq:15}
\end{align}
From Eqs.~\eqref{BSWF} to~\eqref{eq:15}, one arrives at
\begin{align}
\expval{x^m}_{\psi_{0^-}} = \int_0^1 d\alpha \; \alpha^m \Bigg[ \frac{12}{N_{0^-}} \frac{{\cal G}_{0^-}^\nu(\alpha,p_\perp^2)}{\nu+1} \Bigg] \,,
\end{align}
with 
\begin{align}
&
\mathcal{G}_{0^-}^\nu(\alpha,p_\perp^2) = 2^\nu (\nu+1) (\alpha M_{\bar Q} + (1-\alpha)M_q)\nonumber \\
&
\times \left[ \int_{-1}^{1-2\alpha}dw \left( \frac{\alpha}{1- w}\right)^\nu + \int_{1-2\alpha}^1dw \left(\frac{1- \alpha}{1-w}\right)^\nu \right] \nonumber \\
&
\times \frac{\tilde{\rho}_{0^-}^\nu (w)}{\left[p_\perp^2 + \Lambda_{1-2\alpha}^2\right]^{\nu+1}} \,.  
\end{align}
Uniqueness of the Mellin moments implies the connection between the Feynman parameter $\alpha$ and the momentum fraction $x$; therefore one can identify the LFWF as
\begin{equation}
\psi_{0^-}(x,p_\perp^2) = \frac{12}{N_{0^-}} \frac{{\cal G}_{0^-}^\nu(x,p_\perp^2)}{\nu+1} \,.
\end{equation}

\subsection{Parton Distribution Amplitude}
\label{subsec:PDA}

With the LFWF at hand, $\psi_{0^-} (x,p_\perp^2)$, its integration over $p_\perp^2$-dependence yields the PDA:
\begin{equation}
f_{0^-} \phi_{0^-}(x) = \int \frac{d^2p_\perp}{16\pi^3} \, \psi_{0^-}(x,p_\perp^2) \,,
\end{equation}
where $f_{0^-}$ is the leptonic decay constant. The only term in
the above equation that depends on $p_\perp$ is the denominator $(p_\perp^2 + \Lambda_{1-2x}^2)^{\nu+1}$ in ${\cal G}_{0^-}^\nu(x,p_\perp^2)$; therefore, we arrive at the following algebraic relation between the LFWF and the PDA: 
\begin{equation}
\label{eq:LFWFPDArel}
\psi_{0^-}(x,p_\perp^2) = 16\pi^2 f_{0^-} \, \frac{\nu \Lambda_{1-2x}^{2\nu}}{\left(p_\perp^2 + \Lambda_{1-2x}^2\right)^{\nu+1}} \, \phi_{0^-}(x) \,.
\end{equation}
This result is a merit of the algebraic model. Note also that, throughout this manuscript, we shall employ dimensionless and unit normalized PDAs, \emph{i.e.} $\int_0^1\; dx\, \phi_{0^-}(x) = 1$.

It is worth noting here that the degree of $(x,p_\perp^2)$-factorization of the LFWF of a pseudoscalar meson is exposed within our algebraic model through Eqs.~\eqref{eq:Lambda2w} and~\eqref{eq:LFWFPDArel}. In the chiral limit, $m_{0^-}=0$, and quark-antiquark flavor symmetry, $M_{q}=M_{\bar{Q}}$, one has
\begin{equation}
\label{eq:LFWFfact}
\psi_{0^-}^{\text{chiral}}(x,p_\perp^2) =  \left[16\pi^2f_{0^-} \frac{\nu M_q^{2\nu}}{(p_\perp^2+M_q^2)^{\nu+1}}\right] \phi_{0^-}(x) \,.
\end{equation}
where the bracketed term no longer depends on $x$; hence, the $x$ and $p_\perp^2$ dependence of the LFWF is completely factorized in the chiral limit. Contrary, as captured by Eqs.~\eqref{eq:Lambda2w} and~\eqref{eq:LFWFPDArel}, a non-zero meson mass and quark-antiquark flavor asymmetry, \emph{i.e.} $m_{0^-}^2 \neq 0$ and $M_q \neq M_{\bar{Q}}$, yield a LFWF which correlates $x$ and $p_\perp^2$. Therefore, one should expect an increasingly dominant role of $x$ and $p_\perp^2$ correlations in heavy-light systems.

\subsection{Phenomenology of LFWFs from PDAs}
\label{subsec:Results1}

We now have all the necessary tools to compute the LFWF in terms of the PDA within this algebraic model. We proceed then to compute the LFWFs of the lowest-lying heavy-light pseudoscalar mesons, \emph{i.e.} the $D$, $D_s$, $B$, $B_s$ and $B_c$ mesons, from previous determinations of their PDAs. Whilst much has recently been learned about the point-wise behavior of light-meson PDAs~\cite{Chang:2013pq, Chang:2013nia, Segovia:2013eca, Gao:2014bca, Braun:2015axa, Shi:2015esa, Raya:2015gva, Li:2016dzv, Li:2016mah, Gao:2016jka, Chang:2016ouf, Zhang:2017bzy, Gao:2017mmp, Zhang:2017zfe, Chen:2018rwz, Ding:2018xwy}, the information about heavy-light meson DAs remains sketchy~\cite{Braun:2003wx, Lee:2005gza, Grozin:2005iz, Beneke:2011nf, Braun:2012kp, Bell:2013tfa, Beneke:2018wjp}. Up to our knowledge, there are in the literature only two comprehensive analyses of PDAs for all lowest-lying heavy-light pseudoscalar mesons~\cite{Binosi:2018rht, Serna:2020txe}, both based on a symmetry-preserving approach to the continuum bound-state problem in quantum ﬁeld theory.

The PDAs that we use in this manuscript are those computed within the so-called Continuum Schwinger Function Method (CSM)~\cite{Binosi:2018rht}, which are fully compatible with the results obtained in Ref.~\cite{Serna:2020txe}. 

The continuum bound-state problem within CSM is defined by a set of coupled integral equations~\cite{Roberts:1994dr, Maris:2003vk, Eichmann:2016yit} and thus a manageable solution is only obtained once a truncation scheme is specified. References~\cite{Munczek:1994zz, Bender:1996bb} describe a systematic, symmetry-preserving approach whose leading-order term is the widely-used rainbow-ladder (RL) truncation. It is accurate for ground-state light-quark vector- and isospin-nonzero-pseudoscalar-mesons, for related ground-state octet and decouplet baryons~\cite{Chang:2011vu, Bashir:2012fs, Roberts:2015lja, Horn:2016rip, Eichmann:2016yit, Chen:2019fzn}, and also for similar heavy hadrons when a judicious modification is performed~\cite{Ding:2015rkn, Yin:2019bxe, Yin:2021uom}. The RL truncation is precise in these channels because higher-order corrections largely cancel owing to preservation of relevant Ward-Green-Takahashi identities~\cite{Ward:1950xp, Green:1953te, Takahashi:1957xn}. The RL truncation becomes a poor approximation in systems constituted by valence-quarks whose masses are quite different. This is because the mentioned higher-order corrections cannot be effectively canceled in such cases. Truncations which improve upon RL are known~\cite{Chang:2009zb, Bashir:2011dp, Williams:2015cvx, Binosi:2016rxz, Qin:2016fbu, Qin:2020jig, Binosi:2016wcx}, but they have not been tested in heavy-light systems.

\begin{figure}[!t]
\centerline{\includegraphics[width=0.45\textwidth]{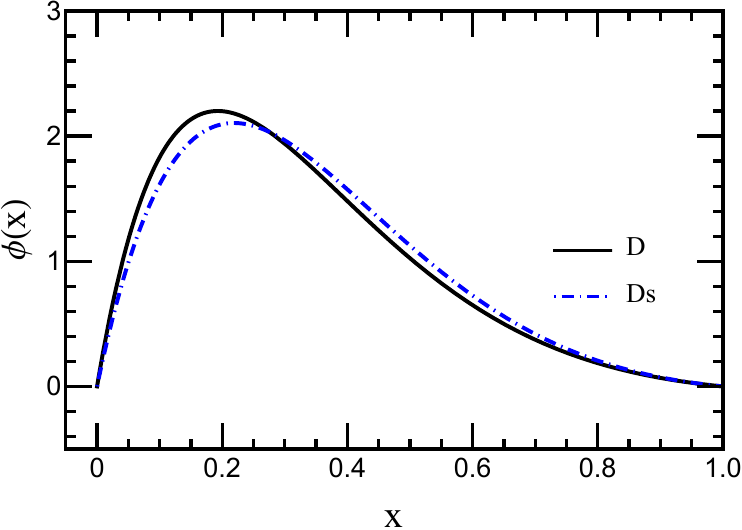}}
\vspace*{0.50cm}
\centerline{\includegraphics[width=0.45\textwidth]{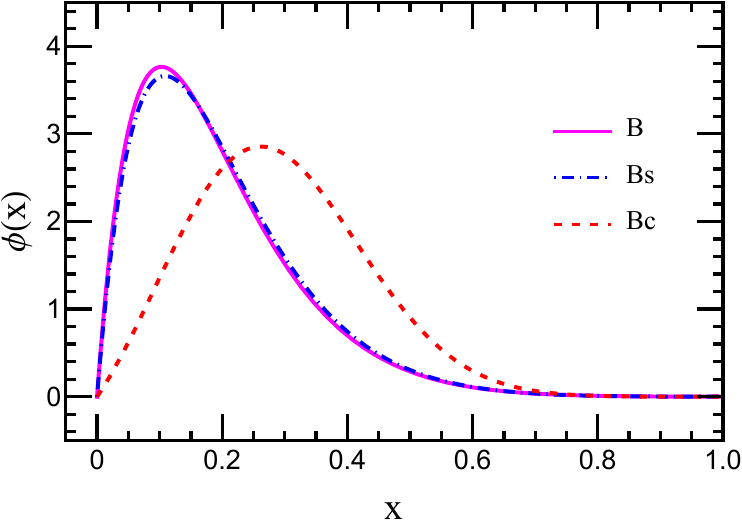}}
\caption{\label{fig:PDAs} Distribution amplitudes of heavy-light mesons. The shaded band surrounding a given curve show the uncertainty in the
associated values of $(\alpha,\beta)$ listed in Table~\ref{tab:AlphaBeta}.}
\end{figure}

Authors of Ref.~\cite{Binosi:2018rht} compute all necessary quantities to obtain the PDAs using the RL truncation on a conservative domain where it is known that the truncation works well, and then extrapolate into the complementary domain using the Schlessinger point method (SPM), whose properties and accuracy are explained elsewhere~\cite{Schlessinger:1966zz, PhysRev.167.1411, Tripolt:2016cya, Chen:2018nsg, Cui:2020rmu}. Distribution amplitudes for $D$, $D_s$, $B$, $B_s$ and $B_c$ mesons are then delivered by Ref.~\cite{Binosi:2018rht} with the following parametrization:
\begin{equation}
\phi_{0^-}(x) = 4 N_{\alpha\beta} x \bar x \, e^{4\alpha^2x\bar x - \beta^2 (x-\bar x)} \,,
\label{eq:PDApara}
\end{equation}
where the parameters $\alpha$ and $\beta$ are determined by requiring a least-squares best-ﬁt to $\expval{x^m}_{\phi_{0^-}}$, with $m=1,2,3$ and
\begin{equation}
\expval{x^m}_{\phi_{0^-}} = \int_0^1 dx \; x^m \, \phi_{0^-}(x) \,.
\end{equation}
Besides, $\bar x =1-x$ and $N_{\alpha\beta}$ is a normalization factor that ensures $\expval{x^0}_{\phi_{0^-}}=1$. The $(\alpha, \beta)$ values are listed in Table~\ref{tab:AlphaBeta} and the associated PDAs are depicted in Fig.~\ref{fig:PDAs}. One can observe that the PDAs become more asymmetric and sharply peaked as the difference between the masses of the meson's valence-quarks increases.

\begin{table}[!t]
\caption{\label{tab:AlphaBeta} The $(\alpha,\beta)$-pairs that specify the PDAs of heavy-light mesons via Eq.~\eqref{eq:PDApara}.}
\begin{ruledtabular}
\begin{tabular}{cccccc}
& $D$ & $D_s$ & $B$ & $B_s$ & $B_c$ \\
\hline
$\alpha$ & $0.265(30)$ & $0.508(30)$ & $0.497(70)$ & $0.669(60)$ & $1.901(70)$ \\
$\beta$  & $1.435(30)$ & $1.391(30)$ & $2.166(60)$ & $2.177(60)$ & $2.163(60)$ \\
\end{tabular}
\end{ruledtabular}
\end{table}

At this point, it is worth mentioning that the quark and antiquark PDAs are connected via momentum conservation through the equality:
\begin{eqnarray}
\label{eq:PDAquarkantiquark}
\phi_{0^-}^q(x)=\phi_{0^-}^{\bar Q}(\bar x) \;,
\end{eqnarray}
which is a constricted and firm relation that prevails even after scale evolution~\cite{Lepage:1979zb, Efremov:1979qk, Lepage:1980fj}. 

It is also important to note that the presented algebraic model is equally valid for any momentum-scale that characterizes the exclusive scattering process in which the meson is involved. That is to say, given the valence-quark (twist-two) PDA of a pseudoscalar heavy-light meson at a given momentum scale, $\zeta$, the formalism provides us the corresponding leading-twist LFWF of the vector meson at the same energy scale. The PDAs, and thus their related LFWFs, are expressed in a quasiparticle basis at an intrinsic scale, $\zeta_H$, for which the valence degrees of freedom fully express the properties of the hadron under study. Most results herein are quoted at $\zeta_H=2\,\text{GeV}$ (unless specified otherwise); however, for the sake of simplicity, the label $\zeta_H$ shall be omitted.

Let us now proceed with the computation of the LFWFs for the $D$, $D_s$, $B$, $B_s$ and $B_c$ mesons through their connection with the corresponding PDAs, Eq.~\eqref{eq:LFWFPDArel}. Besides, the parameters needed to compute the LFWFs are shown in Table~\ref{tab:Parameters}. Note that these are not free-parameters of this analysis, these are computed in connection with the PDAs collected herein and originally reported in Ref.~\cite{Binosi:2018rht}. As shown in Table~\ref{tab:Parameters}, the static properties of pseudoscalar heavy-light mesons are in fairly good agreement with the experimental data collected in the Review of Particle Physics by the Particle Data Group~\cite{ParticleDataGroup:2022pth} and also with the latest results reported by Lattice-QCD~\cite{Chiu:2007bc}.

\begin{table}[!t]
\caption{\label{tab:Parameters} Quoted in GeV, Masses, $M$, and decay constants, $f$, of the lowest-lying heavy-light pseudoscalar mesons reported in Ref.~\cite{Binosi:2018rht}; compared with averages of available experimental and lattice-QCD determinations reported in Refs.~\cite{ParticleDataGroup:2022pth, Chiu:2007bc}. The current-quark masses are $m_{u/d}(\xi=\xi_2=2\,\text{GeV})=0.0049\,\text{GeV}$, $m_{s}(\xi_2)=0.114\,\text{GeV}$, $m_c(\xi_2)=1.25\,\text{GeV}$, $m_b(\xi_2)=4.35\,\text{GeV}$ which correspond to the following dressed quark masses $M_{u/d}(\xi=\xi_0=0\,\text{GeV})=0.533\,\text{GeV}$, $M_s(\xi_0)=0.717\,\text{GeV}$, $M_c(\xi_0)=1.558\,\text{GeV}$, $M_b(\xi_0)=4.651\,\text{GeV}$.}
\begin{ruledtabular}
\begin{tabular}{lcccccc}
& \multicolumn{2}{c}{Theory} & \multicolumn{2}{c}{Experiment} & \multicolumn{2}{c}{Lattice} \\
& $M$ & $f$ & $M$ & $f$ & $M$ & $f$ \\
\hline
$D$   & $1.88(5)$  & $0.158(8)$  & $1.87$ & $0.153(7)$  & $1.87$    & $0.150(1)$ \\
$D_s$ & $1.94(4)$  & $0.171(6)$  & $1.97$ & $0.177(3)$  & $1.97$    & $0.176(1)$ \\
$B$   & $5.30(15)$ & $0.142(13)$ & $5.28$ & $0.138(19)$ & $5.28$    & $0.132(3)$ \\
$B_s$ & $5.38(13)$ & $0.179(12)$ & $5.37$ & $\cdots$    & $5.37$    & $0.161(2)$ \\
$B_c$ & $6.31(1)$  & $0.367(1)$  & $6.27$ & $\cdots$    & $6.28(1)$ & $0.346(3)$ \\
\end{tabular}
\end{ruledtabular}
\end{table}


\begin{figure}[!t]
\centerline{\includegraphics[width=0.45\textwidth]{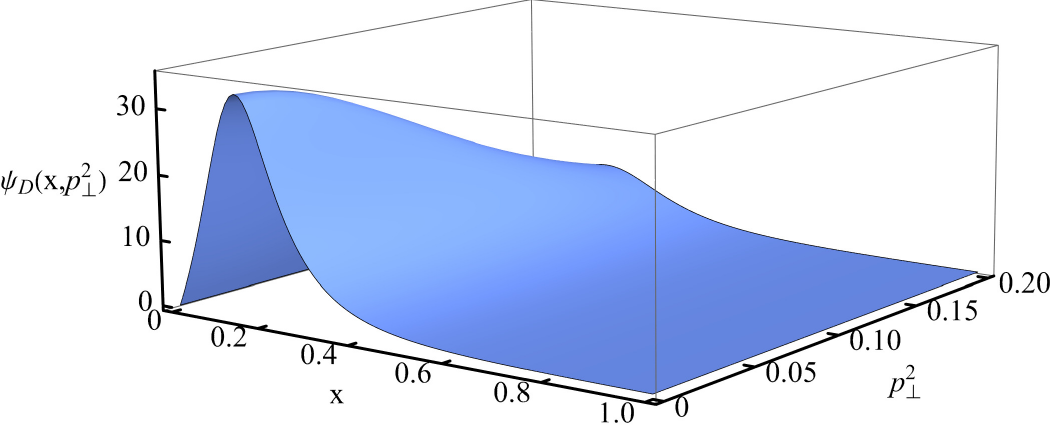}}
\vspace*{0.60cm}
\centerline{\includegraphics[width=0.45\textwidth]{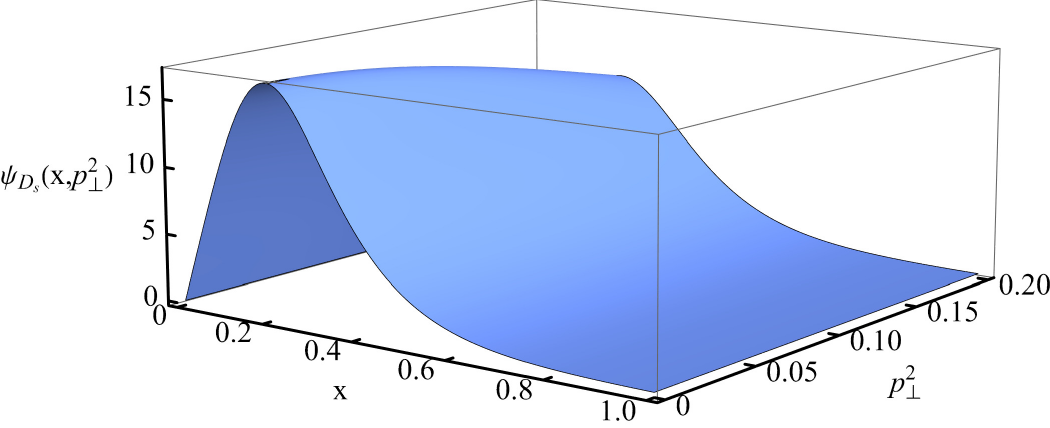}}
\caption{\label{fig:LFWFsDmesons} LFWFs of the $D$-meson (upper panel) and $D_s$ one (lower panel). As one can guess from Eq.~\eqref{eq:LFWFPDArel}, the LFWF has mass dimension $-1$, which is expressed in GeV$^{-1}$; whereas $p_\perp^2$ has dimension GeV$^2$ and $x$ is dimensionless.}
\end{figure}

Figure~\ref{fig:LFWFsDmesons} shows the leading-twist LFWFs of the lowest-lying pseudo-scalar charmed mesons. It is evident that the LFWF of the $D$-meson exhibits a more pronounced $x$-dependence, being narrower than the one corresponding to the $D_s$-meson. Moreover, both LFWFs indicate that the most probable fraction of light-front total momentum carried by the light quark inside the charmed meson is around $x=0.2$. The LFWF of the $D$-meson presents a strong $p_\perp^2$-dependence; for instance, the maximum in $x$ is one sixth of its original value when modifying $p_\perp^2$ from $0.0$ to $0.2$ GeV$^2$. The fall-off along the $p_\perp^2$-range depicted is smoother for the LFWF of the $D_s$-meson. Its extreme in the variable $x$ decreases towards two thirds of its original value at $p_\perp^2=0.2\,\text{GeV}^2$. It is worth emphasizing that the maximum in the LFWFs does not move in $x$ as varying the value of $p_\perp^2$. The features presented can be directly connected with the masses of the valence dressed quarks that constitute the hadron.

\begin{figure}[!t]
\centerline{\includegraphics[width=0.45\textwidth]{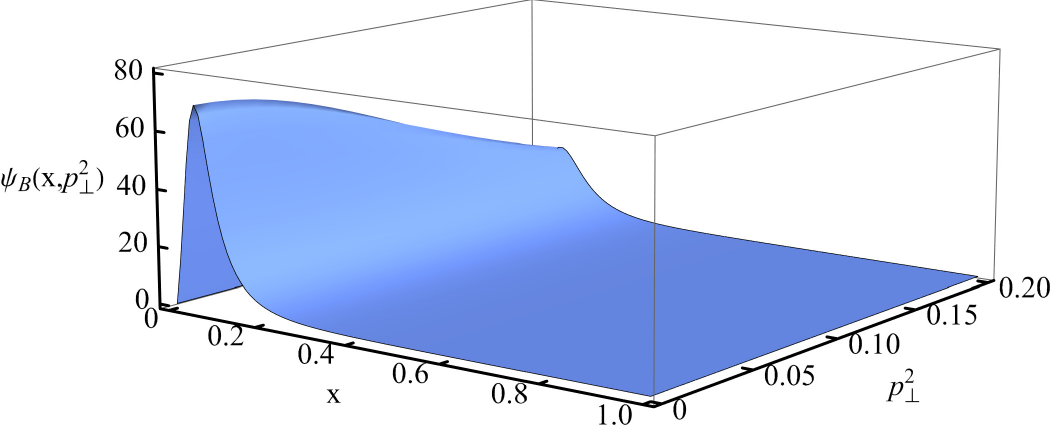}}
\vspace*{0.60cm}
\centerline{\includegraphics[width=0.45\textwidth]{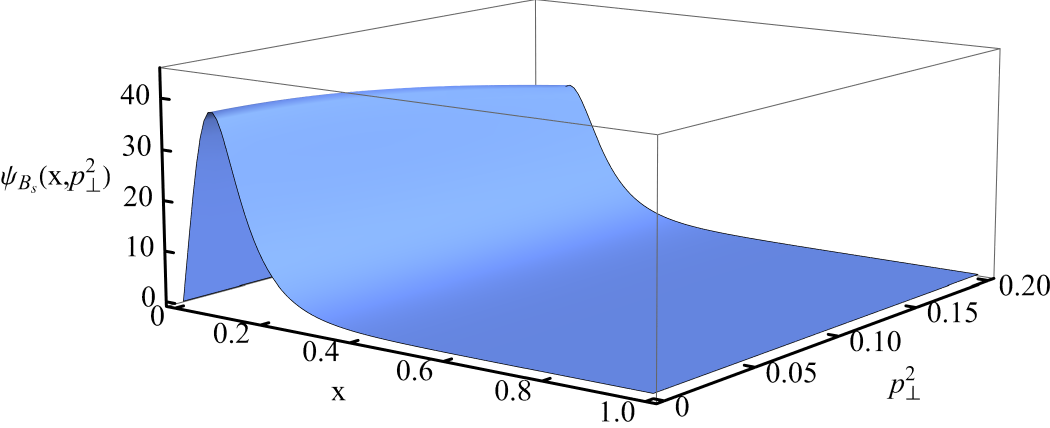}}
\vspace*{0.60cm}
\centerline{\includegraphics[width=0.45\textwidth]{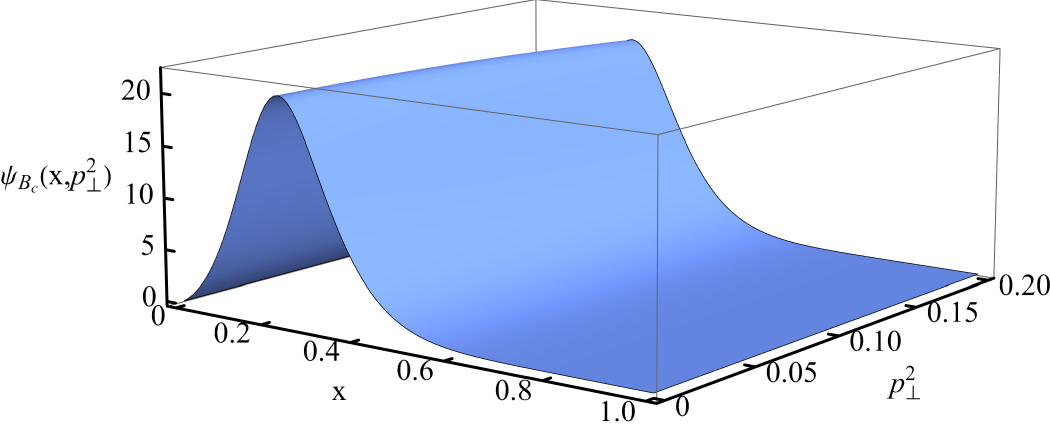}}
\caption{\label{fig:LFWFsBmesons} LFWFs of the $B$-meson (upper panel), $B_s$-meson (middle panel) and $B_c$-meson (lower panel). As one can guess from Eq.~\eqref{eq:LFWFPDArel}, the LFWF has mass dimension $-1$, which is expressed in GeV$^{-1}$; whereas $p_\perp^2$ has dimension GeV$^2$ and $x$ is dimensionless.}
\end{figure}

Figure~\ref{fig:LFWFsBmesons} shows the leading-twist LFWFs of the lowest-lying pseudo-scalar bottom mesons: $B$-meson (upper panel), $B_s$-meson (middle panel) and $B_c$-meson (lower panel). A clear pattern is shown, the $x$ point-wise behavior of the LFWF is the narrowest for the $B$-meson and it becomes successively wider for the $B_s$ and $B_c$ mesons. Moreover, at $p_\perp^2=0$, the LFWF peaks at $x=0.8$, $0.10$ and $0.25$ for the $B$, $B_s$ and $B_c$, respectively. Besides, the $B_c$'s LFWF is more symmetric than the other two. With respect to the $p_\perp^2$-dependence, the LFWF of the $B$-meson exhibits the strongest behavior, with its maximum decaying two thirds of its original value at $p_\perp^2=0.2\,\text{GeV}$. The LFWFs of the $B_s$ and $B_c$ mesons are much smoother in $p_\perp^2$-dependence, the maximum of the $B_s$ LFWF is still three fourths of its initial value at $p_\perp^2=0.2\,\text{GeV}$, whereas the LFWF of the $B_c$ meson is almost constant along the $p_\perp^2$-range depicted. Again, one can mention that the maximum in the LFWFs does not move in $x$ as varying the value of $p_\perp^2$ and that all the properties highlighted here can be traced back to the masses of the valence quarks in the meson.  

\begin{table*}[!t]
\caption{\label{tab:MM-LFWFs} Computed $\expval{(2x-1)^m}$ moments, with $m=0$, $1$, $2$, $\ldots$, $10$, at $p_\perp^2=0.0$, $0.1$ and $0.2\,\text{GeV}^2$, of the leading-twist LFWFs of the lowest-lying pseudo-scalar charmed and bottom mesons. All quantities are given in GeV$^{-1}$.}
\begin{ruledtabular}
\scalebox{0.90}{
\begin{tabular}{cc|ccccccccccc}
\multicolumn{2}{c|}{$\expval{(2x-1)^m}$} & $m=0$ & 1 & 2 & 3 & 4 & 5 & 6 & 7 & 8 & 9 & 10 \\
\hline
$D$ & $p_\perp^2=0.0$ & 6.82 & -4.24 & 2.97 & -2.15 & 1.62 & -1.25 & 0.99 & -0.79 & 0.65 & -0.54 & 0.45 \\
    & $p_\perp^2=0.1$ & 5.07 & -3.03 & 2.10 & -1.51 & 1.14 & -0.88 & 0.70 & -0.56 & 0.46 & -0.38 & 0.32 \\
    & $p_\perp^2=0.2$ & 2.72 & -1.47 & 0.99 & -0.69 & 0.52 & -0.40 & 0.32 & -0.26 & 0.21 & -0.18 & 0.15 \\[2ex]
$D_s$ & $p_\perp^2=0.0$ & 5.49 & -2.64 & 1.72 & -1.14 & 0.84 & -0.62 & 0.49 & -0.38 & 0.31 & -0.26 & 0.22 \\
      & $p_\perp^2=0.1$ & 5.09 & -2.42 & 1.58 & -1.05 & 0.77 & -0.57 & 0.45 & -0.35 & 0.29 & -0.24 & 0.20 \\
      & $p_\perp^2=0.2$ & 4.13 & -1.91 & 1.25 & -0.82 & 0.61 & -0.45 & 0.36 & -0.28 & 0.23 & -0.19 & 0.16 \\[1ex]
\hline
\tstrut
$B$   & $p_\perp^2=0.0$ & 6.28 & -5.24 & 4.46 & -3.84 & 3.35 & -2.94 & 2.60 & -2.31 & 2.07 & -1.85 & 1.67 \\
      & $p_\perp^2=0.1$ & 5.10 & -4.19 & 3.53 & -3.02 & 2.61 & -2.27 & 2.00 & -1.77 & 1.58 & -1.41 & 1.27 \\
      & $p_\perp^2=0.2$ & 3.23 & -2.56 & 2.10 & -1.75 & 1.48 & -1.27 & 1.10 & -0.96 & 0.85 & -0.75 & 0.67 \\[2ex]
$B_s$ & $p_\perp^2=0.0$ & 5.38 & -4.13 & 3.30 & -2.69 & 2.24 & -1.88 & 1.61 & -1.38 & 1.20 & -1.05 & 0.93 \\
      & $p_\perp^2=0.1$ & 5.08 & -3.88 & 3.08 & -2.51 & 2.08 & -1.75 & 1.49 & -1.28 & 1.11 & -0.98 & 0.86 \\
      & $p_\perp^2=0.2$ & 4.30 & -3.25 & 2.56 & -2.07 & 1.71 & -1.43 & 1.22 & -1.04 & 0.90 & -0.79 & 0.69 \\[2ex]
$B_c$ & $p_\perp^2=0.0$ & 5.10 & -2.53 & 1.47 & -0.92 & 0.61 & -0.42 & 0.30 & -0.23 & 0.17 & -0.13 & 0.11 \\
      & $p_\perp^2=0.1$ & 5.02 & -2.49 & 1.45 & -0.90 & 0.60 & -0.42 & 0.30 & -0.22 & 0.17 & -0.13 & 0.11 \\
      & $p_\perp^2=0.2$ & 4.79 & -2.37 & 1.38 & -0.86 & 0.57 & -0.40 & 0.29 & -0.21 & 0.16 & -0.13 & 0.10 \\
\end{tabular}
}
\end{ruledtabular}
\end{table*}

For completeness, Table~\ref{tab:MM-LFWFs} shows the computed $\expval{(2x-1)^m}$ moments, with $m=0$, $1$, $2$, $\ldots$, $10$, at $p_\perp^2=0.0$, $0.1$ and $0.2$ GeV$^2$, of the leading-twist LFWFs of the lowest-lying pseudo-scalar charmed and bottom mesons. The most salient features of this analysis are: (i) even moments are positive and odd ones are negative but, in absolute value, they systematically fall-off towards zero being always the former larger than the later; (ii) higher order moments have lower values for all mesons being in general the last reported moment an order of magnitude smaller than the first one; (iii) the value of a given moment decreases as $p_\perp^2$ increases for any meson, however, once such a value is small enough it remains nearly constant with respect to changes in $p_\perp^2$; and (iv) the symmetric shape in the $x$-dependence of the $B_c$ LFWF can be deduced from the rapid fall-off of high order moments.


\section{Internal structure of Pseudoscalar Heavy-light mesons}
\label{sec:IntStrMes}

\subsection{Generalized Parton Distribution Function}
\label{subsec:GPDs}

Generalized parton distributions (GPDs) were introduced in the factorization analysis of some hard exclusive processes by a series of authors~\cite{Muller:1994ses, Radyushkin:1996nd, Ji:1996nm, Radyushkin:1996ru, Radyushkin:1997ki, Ji:1996ek}. GPDs are formally defined as off-diagonal hadronic expectation values of non-local quark- or gluon-field operators projected onto the light-front. Therefore, a naive interpretation relates the GPDs with transitions between hadron states that involve momentum and/or helicity transfers and thus they encode a large amount of information about the nature of these reactions and the structure of involved hadrons.

GPDs are difficult to calculate from QCD's first principles. However, their properties, which can be deduced from the fundamental features and symmetries of QCD, serve to develop modeling strategies. In fact, the most common approaches to the evaluation of GPDs are the so-called double distribution~\cite{Radyushkin:1997ki} and overlap~\cite{Diehl:2003ny} representations. While the double distribution approach fulfills, at the price of hiding positivity, the property of polynomiality inherited from the invariance of QCD's action under Lorentz transformations, the overlap representation complies with the positivity condition imposed by the underlying Hilbert-space's norm but spoils a practical management of polynomiality. 

The covariant extension approach~\cite{Mezrag:2014jka, Chouika:2017rzs, Chouika:2017dhe, Chavez:2021llq, Chavez:2021koz} has recently been developed with the main goal of building GPDs that fulfill by construction with the two fundamental properties mentioned above, positivity and polynomiality. The recipe of the covariant extension\footnote{The interested reader is referred to Refs.~\cite{Chouika:2018mbk, Mezrag:2022pqk} for further details.} needs as a first step the construction of positive DGLAP GPDs through the overlap representation of light-front wave-functions. Therefore, the algebraic model appears as a very suitable framework to perform such a step in the case of pseudo-scalar heavy-light mesons. The valence quark GPD can be obtained as
\begin{equation}
H_{0^-}(x,\xi,t) = \int \frac{d^2p_{\perp}}{16\pi^3} \, \psi_{0^-}^{\ast} \left( x^-, (\mathbf{p}_\perp^{-})^2 \right) \psi_{0^-}  \left( x^+, (\mathbf{p}^{+}_\perp)^2 \right) \,, \\
\label{GPDdefinition}
\end{equation}
where
\begin{equation}
x^{\pm} = \frac{x\pm \xi}{1\pm \xi} \,, \hspace*{0.60cm} \mathbf{p}_\perp^\pm = \mathbf{p}_\perp \mp \frac{\mathbf{\Delta}_\perp}{2}\frac{1-x}{1\pm \xi}\;.
\end{equation}
If $p\,(p')$ denotes the initial (final) meson momentum, then $P=(p+p')/2$ is the total momentum of the system and $t=-\Delta^2=-(p-p')^2$ is the momentum transfer; $\Delta_\perp^2 = \Delta^2(1-\xi^2) -4\xi^2 m_{\text{M}}^2$. In addition, the longitudinal momentum fraction transfer is  $\xi=[-n\cdot \Delta]/[2n\cdot P]$.  Both $x$ and $\xi$ have support on $[-1,1]$, but the overlap representation is only considered in the DGLAP region, $|x| > |\xi|$. Its kinematical completion to the ERBL domain can be achieved through the covariant extension~\cite{Chouika:2017dhe, Chouika:2017rzs, Chavez:2021koz, Chavez:2021llq}, up to the named $D$-terms and so it is a nontrivial task which goes beyond the scope of this manuscript; hence, in the following, we shall restrain ourselves to $\xi \geq 0$ and $|x| > |\xi|$. Note once more that Eq.~\eqref{GPDdefinition} implies that the meson is described as a quark-antiquark Fock state. This picture is then valid at the hadronic scale, in which the fully dressed quark/antiquark quasiparticles encode all the properties of the meson. 

The valence quark GPD can be worked out in detail by substituting Eq.~\eqref{eq:LFWFPDArel} into Eq.~\eqref{GPDdefinition}:
\begin{align}
&
H_{0^-}(x,\xi,t) = (16\pi^2 f_{0^-}\nu)^2 \phi_{0^-}(x^+)\phi_{0^-}(x^-) \Lambda_{1-2x^+}^{2\nu} \Lambda_{1-2x^-}^{2\nu} \nonumber \\
&
\times \int \frac{d^2p_\perp}{16\pi^3} \frac{1}{((\bold{p}_\perp^-)^2+\Lambda_{1-2x^-}^2)^{\nu+1}}\frac{1}{((\bold{p}_\perp^+)^2+\Lambda_{1-2x^+}^2)^{\nu+1}} \,,
\label{eq:GPDint1}
\end{align}
in such a way that performing a Feynman parametrization and after a suitable change of variables, it turns into
\begin{align}
&
\int \frac{d^2p_\perp}{16\pi^3} \frac{1}{((\bold{p}_\perp^-)^2+\Lambda_{1-2x^-}^2)^{\nu+1}}\frac{1}{((\bold{p}_\perp^+)^2+\Lambda_{1-2x^+}^2)^{\nu+1}} \nonumber \\
&
= \frac{2\pi}{16\pi^3} \frac{\Gamma(2\nu+2)}{\Gamma^2(\nu+1)} \nonumber \\
&
\times \int_0^1 du u^\nu(1-u)^\nu \int_0^\infty dk_\perp \frac{k_\perp}{(k_\perp^2+\mathbb{M}^2(u))^{2\nu+2}} \nonumber \\
&
= \frac{1}{16\pi^2}\frac{\Gamma(2\nu+2)}{\Gamma^2(\nu+1)}\int_0^1 du \frac{u^\nu (1-u)^\nu}{[\mathbb{M}^2(u)]^{2\nu+1}} \,,
\label{eq:GPDint2}
\end{align}
where the function $\mathbb{M}^2(u)$ depends on the model parameters, as well as the kinematic variables $x$, $\xi$ and $t$. It acquires the form $\mathbb{M}^2(u)=c_2 u^2 + c_1 u + c_0$, with
\begin{subequations}
\begin{align}
c_2 &= \frac{(1-x)^2}{(1-\xi^2)^2}t \,, \\
c_1 &= -\frac{(1-x)^2}{(1-\xi^2)^2}t + \Lambda^2_{1-2x^+}-\Lambda^2_{1-2x^-} \,, \\
c_0 &= \Lambda^2_{1-2x^-} \,.
\end{align}
\end{subequations}
Therefore, the final expression for the valence quark GPD is given by
\begin{align}
H_{0^-}(x,\xi,t) &= (f_{0^-}\nu)^2 \phi_{0^-}(x^+) \phi_{0^-}(x^-) \Lambda_{1-2x^+}^{2\nu} \Lambda_{1-2x^-}^{2\nu} \nonumber \\
&
\times \frac{\Gamma(2\nu+2)}{\Gamma^2(\nu+1)}\int_0^1 du \frac{u^\nu (1-u)^\nu}{[\mathbb{M}^2(u)]^{2\nu+1}} \,.
\label{eq:GPDfinal}
\end{align}

Note that the function $\mathbb{M}^2(u)$ reduces in the chiral limit to
\begin{equation}
\mathbb{M}^2(u)= -t\,u(1-u)\frac{(1-x)^2}{(1-\xi^2)^2}+M_q^2\;,
\end{equation}
and so the integration in Eq.~\eqref{eq:GPDfinal} can be carried out algebraically for specific values of $\nu\,\textgreater -1$. In particular, $\nu = 1$ recovers the results in~\cite{Chouika:2017rzs, Mezrag:2016hnp, Mezrag:2014jka, Chavez:2021koz, Chavez:2021llq}. Beyond the chiral limit, an algebraic expression is found for $t = 0$:
\begin{align}
H_{0^-}&(x,\xi,0) = \mathcal{N} \phi_{0^-}^q(x^+)\phi_{0^-}^q(x^-) \frac{\Lambda_{1-2x^+}^{2\nu}}{\Lambda_{1-2x^-}^{2\nu}}  \frac{\Gamma(2\nu+2)}{\Lambda_{1-2x^-}^2} \nonumber \\
& 
\times \;_2\tilde{F}_1\left(1+\nu,1+2\nu,2\nu+2,1-\frac{\Lambda_{1-2x^+}^{2}}{\Lambda_{1-2x^-}^{2}}\right) \,,
\label{eq:GPDxi}
\end{align}
where $_p\tilde{F}_q(u,v,w,z)$ is the regularized hypergeometric function. Conversely, taking $\xi=0$, an expansion of $\mathbb{M}^2(u)$ around $-t\approx 0$ also yields an algebraic solution for Eq.~\eqref{eq:GPDfinal}:
\begin{align}
H_{0^-}(x,0,t) & \overset{t\to 0}{\approx}   \mathcal{N}\frac{[\phi_{0^-}^{q}(x)]^2}{\Lambda_{1-2x}^2} \nonumber \\
&
\times \left[1-c_\nu^{(1)}(1-x)^2 \left( \frac{-t}{\Lambda_{1-2x}^2} \right) + \ldots \right] \,,
\label{eq:exptGPD}
\end{align}
with
\begin{align}
c_\nu^{(1)}= \frac{(1+\nu)(1+2\nu)}{2(3+2\nu)} \,, \hspace*{0.6cm} \mathcal{N}=\left[\int_0^1 dx\; \frac{\phi_{0^-}^2(x)}{\Lambda_{1-2x}^2}  \right]^{-1}\;.
\end{align}

\begin{figure}[!t]
\centerline{\includegraphics[width=0.45\textwidth]{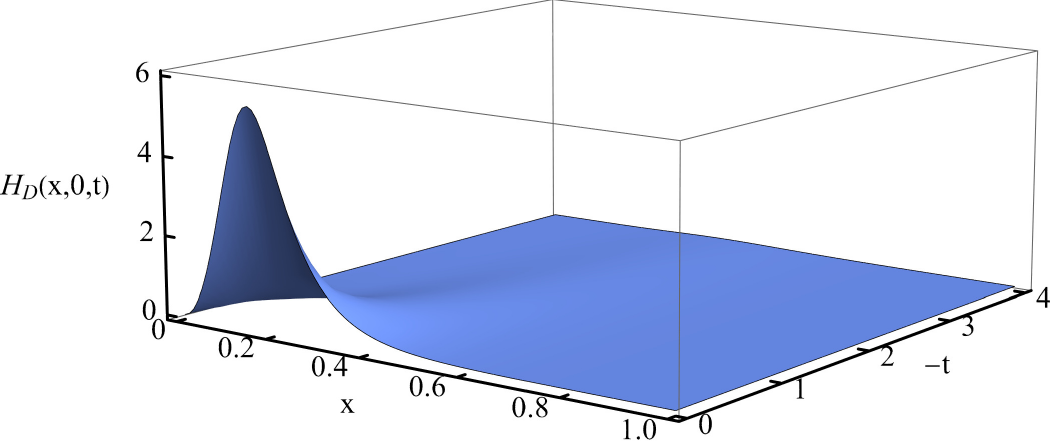}}
\vspace*{0.50cm}
\centerline{\includegraphics[width=0.45\textwidth]{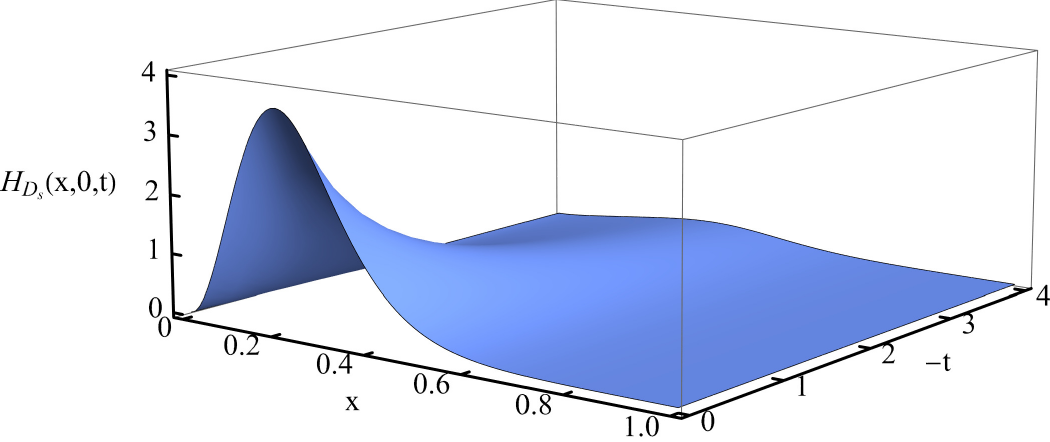}}
\caption{\label{fig:GPDs_Dmesons} Valence quark GPDs obtained from Eq.~(\ref{eq:GPDfinal}) for $\xi=0$. \emph{Upper panel:} $D$-meson GPD and \emph{lower panel:} $D_s$-meson GPD. Mass units in GeV.}
\end{figure}

The $D$- and $D_s$-meson GPDs, for $\xi=0$, are drawn in Figure~\ref{fig:GPDs_Dmesons}. The $D$-meson GPD (upper panel) presents a very sharp behavior with respect to the momentum transfer; in fact, it is almost zero beyond $-t \approx 1\,\text{GeV}^2$. Besides, the $x$-dependence of the $D$-meson GPD is also sharp; weighted at $x\lesssim0.5$, with a maximum at around $x=0.2$ and presenting negligible values for $x\gtrsim0.5$. The $D_s$-meson GPD (lower panel) shows similar but clearly smoother features, \emph{i.e.} it is close to zero for values of momentum transfer larger than $2\,\text{GeV}^2$, the $x$-dependence is weighted on its left for the light quark, with a maximum at around $x=0.25$ and having negligible values for $x\gtrsim0.6$. These properties can be traced back with the dressed-quark and -antiquark content of the analyzed mesons.

\begin{figure}[!t]
\centerline{\includegraphics[width=0.45\textwidth]{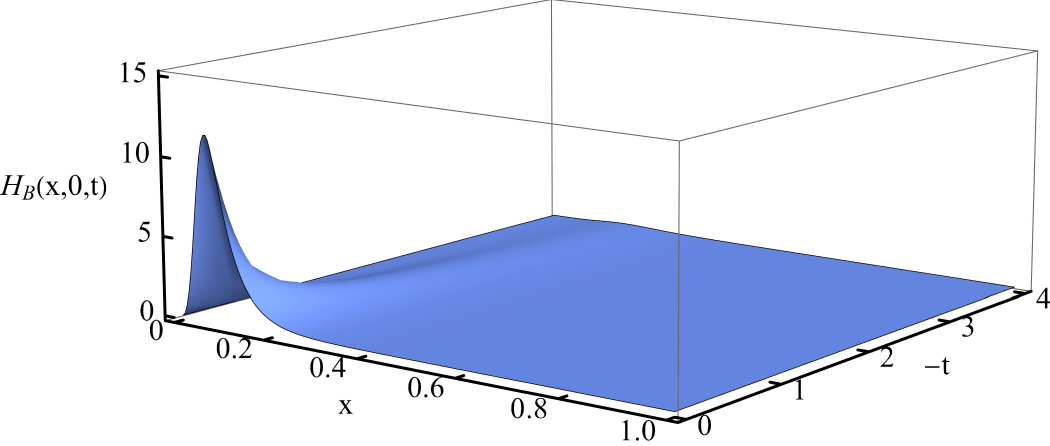}}
\vspace*{0.50cm}
\centerline{\includegraphics[width=0.45\textwidth]{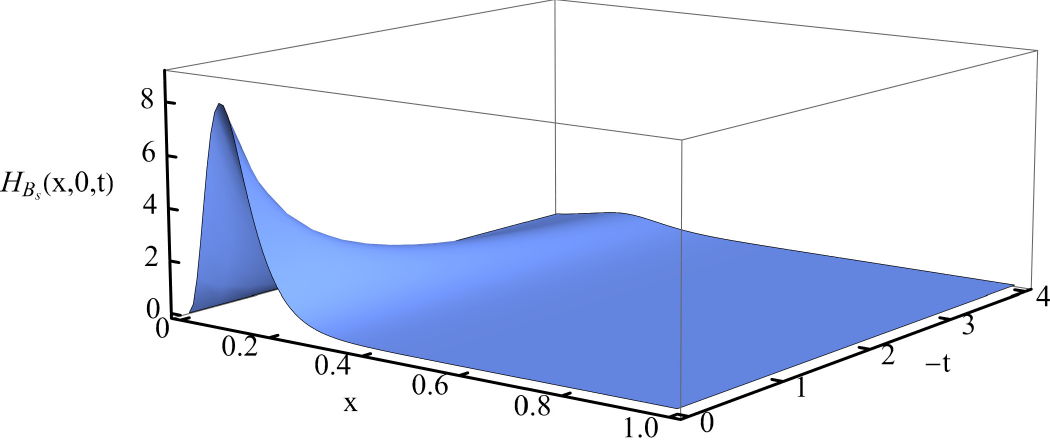}}
\vspace*{0.50cm}
\centerline{\includegraphics[width=0.45\textwidth]{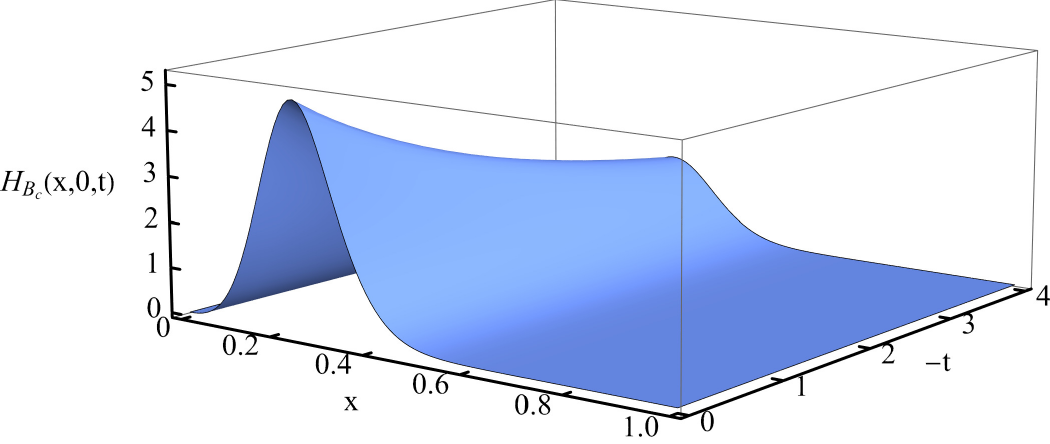}}
\caption{\label{fig:GPDs_Bmesons} Valence quark GPDs obtained from Eq.~(\ref{eq:GPDfinal}) for $\xi=0$. \emph{Upper panel:} $B$-meson GPD, \emph{middle panel:} $B_s$-meson GPD and \emph{lower panel:} $B_c$-meson GPD. Mass units in GeV.}
\end{figure}

Figure~\ref{fig:GPDs_Bmesons} shows the $B$, $B_s$ and $B_c$-meson GPDs for $\xi=0$. In a progressive order, from lighter to heavier mesons, the valence quark GPD shows a strong dependence with respect the transferred momenta, it is relatively narrow in the $x$-range depicted and its maximum moves towards larger values of the meson's longitudinal-momentum fraction. These features are remarkably striking for the $B$ and $B_s$ mesons indicating that they are related not only with the meson's quark-antiquark content but also with the mass difference between the valence dressed-quark and antiquark pair. In the case of the $B_c$-meson, where both quark and antiquark are heavy, the $t$-dependence is the smoothest, with significant nonzero values of the GPD for $-t\gtrsim4\,\text{GeV}^2$; moreover, its point-wise behavior with respect to the fraction of the meson's longitudinal momentum is wider with a maximum more centered within the $x$-range, at $x\approx0.3$.

We now proceed to discuss the derivation of parton distribution functions (PDFs), electromagnetic form factors (EFFs) and Impact parameter space GPDs (IPS-GPDs), as inferred from the knowledge of the GPDs in the DGLAP kinematic region.


\begin{figure*}[!t]
\includegraphics[width=0.49\textwidth]{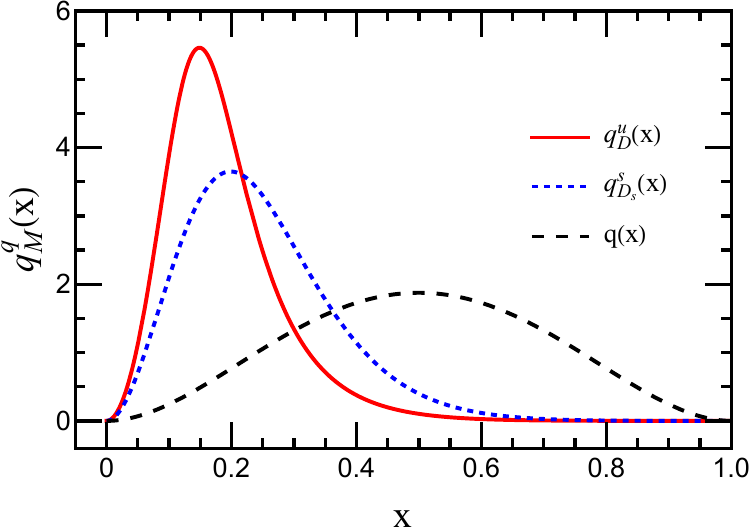}
\includegraphics[width=0.49\textwidth]{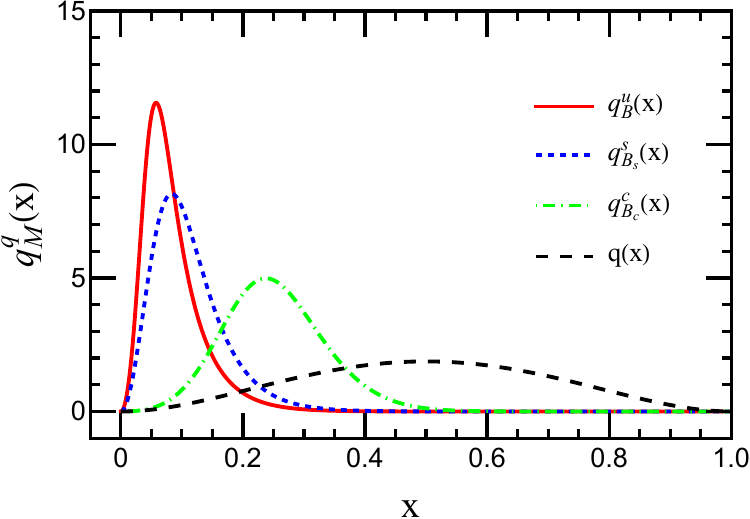}
\caption{\label{fig:PDFs} Valence (light) quark PDFs at the hadron scale, $\zeta_H$. \emph{Left panel:} The solid (red) line corresponds to the $D$-meson case whereas the dotted (blue) line referrers to the $D_s$ one. \emph{Right panel:} The solid (red) line corresponds to the $B$-meson case, the dotted (blue) line reflects the case of $B_s$ state, and the dot-dashed (green) line is related with the $B_c$-meson. For all these panels, the dashed (black) line corresponds to the scale-free parton-like profile $q(x) = 30 x^2 (1-x)^2$.}
\end{figure*}

\subsection{Parton distribution function}

The forward limit of the GPD ($t=0$, $\xi=0$) defines the valence quark PDF:
\begin{equation}
q_{0^-}(x) \equiv H_{0^-}(x,0,0) \,.
\end{equation}
The corresponding antiquark PDF is simply obtained as
\begin{equation}
\label{eq:antiPDF1}
\bar{Q}_{0^-}(\zeta_H;x) = q_{0^-}(\zeta_H;1-x) \,,
\end{equation}
because at the hadron scale, $\zeta_H$, the dressed valence quarks express all hadron properties, in particular, the hadron's momentum is fully carried by them.

Figure~\ref{fig:PDFs} shows the corresponding valence (light) quark PDFs of the lowest-lying pseudo-scalar charmed mesons in the left panel and those of the bottom partners in the right panel. As a reference, the dashed (black) line is the conformal parton-like PDF: $q(x)=30x^2(1-x)^2$. One can see that all drawn PDFs are more pronounced than $q(x)$ and, obviously, asymmetric with respect $q(x)$ as they represent heavy-light mesons. When comparing the PDFs by quark sector, it is clear that the PDF becomes wider and the value of fractional longitudinal momentum moves towards larger values as the light valence-quark is getting heavier. Moreover, the valence light quark PDF of the $B_c$-meson appears quite displaced in $x$ with respect to the PDFs of the $B$ and $B_s$ mesons.

From the experimental point of view, the access and interpretation of PDFs (and GPDs) at $\zeta_H$ imply certain technical and conceptual complications~\cite{Ellis:1996mzs}. In order to produce a consistent picture when comparing experimental data with theoretical predictions, one needs to evolve the hadronic scale PDF. The correct way to do this is by using the all orders scheme introduced in Refs.~\cite{Ding:2019lwe, Ding:2019qlr, Cui:2020tdf, Cui:2020dlm}. It is based upon the assumption that an effective charge $\hat{\alpha}$ allows all beyond leading-order effects to be absorbed within it, thus arriving at a leading-order-like DGLAP evolution equation. There are no experimental data to compare with and therefore this procedure is not necessary at this time.


\begin{figure*}[!t]
\includegraphics[width=0.49\textwidth]{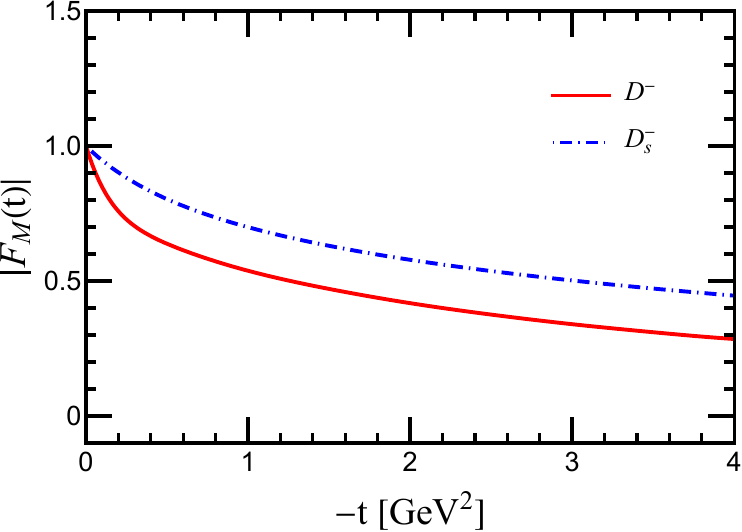}
\includegraphics[width=0.49\textwidth]{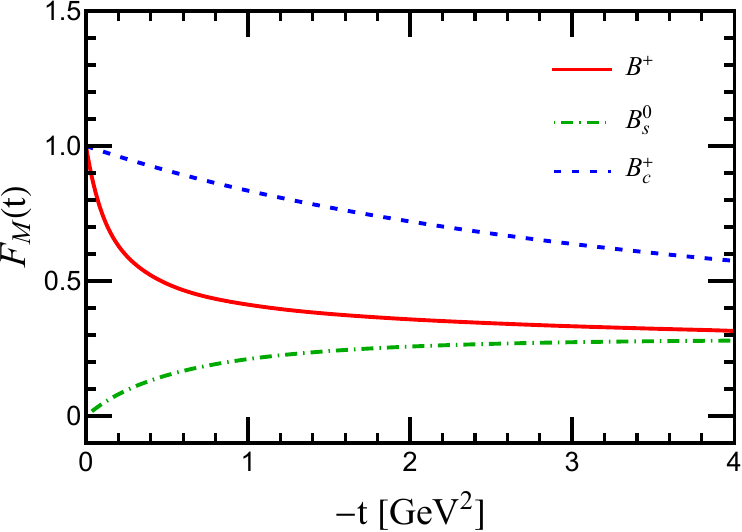}
\caption{\label{fig:EFFs} \emph{Left panel:} Electromagnetic form factors of the lowest-lying pseudo-scalar charmed mesons. \emph{Right panel:} Analogous results for the bottom mesons. We plot the EFF corresponding to charged state whose quark-antiquark content match the $q\bar{Q}$ combination where $q$ is the light quark and $\bar{Q}$ the heavy quark, \emph{i.e.} $D^-=d\bar c$, $D_s^-=s\bar c$, $B^{+}=u\bar b$, $B_s=s\bar b$ and $B_c^+=c\bar b$.}
\end{figure*}

\subsection{Electromagnetic form factor}

The contribution of the $q$-quark to the meson's elastic electromagnetic form factor (EFF) is obtained from the zeroth moment of the GPD,
\begin{equation}
F_{0^-}^q(t) = \int_{-1}^1 dx\; H_{0^-}^q(x,\xi,t) \,,
\label{eq:EFFq}
\end{equation}
an analogous expression holds for the antiquark $\bar{Q}$. Therefore, the complete meson's EFF is given by
\begin{equation}
F_{0^-}(t) = e_q \, F_{0^-}^q(t) + e_{\bar{Q}} \, F_{0^-}^{\bar{Q}}(t) \,, 
\label{eq:MFF1}
\end{equation}
where $e_{q(\bar{Q})}$ is the electric charge of the light quark (heavy antiquark) in units of the positron charge. Due to the polynomiality property of the GPD, the EFF does not depend on $\xi$, therefore one can simply take $\xi=0$. A Taylor expansion around $t\approx 0$ yields
\begin{eqnarray}
\label{eq:chargeradius}
F_{0^-}^q(t) &\overset{t\to 0}{\approx}&1-\frac{(r_{0^-}^q)^2}{6}(-t)+...\;,\\
(r_{0^-}^q)^2 &=&\left. -6 \frac{d F_{0^-}^q(t)}{dt} \right|_{t=0}\;,
\end{eqnarray}
where $r_{0^-}^q$ denotes the contribution of the quark $q$ to the meson charge radius, $r_{0^-}$. Comparing the above equations with the integration on $x$ of Eq.~\eqref{eq:exptGPD}, one obtains a semi-analytical expression for $r_{0^-}^q$:
\begin{equation}
\label{eq:crquark}
(r_{0^-}^q)^2=6 \, \int_0^1 dx\;\hat{f}_{0^-}^q(x)q_{0^-}(x) \,,
\end{equation}
with
\begin{equation}
\hat{f}_{0^-}^q(x) = \frac{c_\nu^{(1)}(1-x)^2}{\Lambda_{1-2x}^2} \,.
\label{eq:fqEFF}
\end{equation}
The above expresions show that the charge radius is tightly connected with the hadronic scale PDF, and thus with the corresponding PDA. The antiquark result is obtained analogously; its contribution to $r_{0^-}$ reads
\begin{equation}
(r_{0^-}^{\bar{Q}})^2=6 \,\int_0^1 dx\;\hat{f}_{0^-}^{\bar{Q}}(x)q_{0^-}(1-x)\,,
\label{eq:crantiquark}
\end{equation}
where $\hat{f}_{0^-}^{\bar{h}}(x)$ is defined in analogy to its quark counterpart in Eq.~\eqref{eq:fqEFF},
\begin{equation}
\hat{f}_{0^-}^{\bar{Q}}(x) = \frac{c_\nu^{(1)}(1-x)^2}{\Lambda_{2x-1}^2}\,.
\end{equation}
Summing up the quark and antiquark contributions, the meson charge radius reads:
\begin{equation}
\label{eq:crgeneral}
r_{0^-}^2= e_q (r_{0^-}^q)^2 + e_{\bar{Q}} (r_{0^-}^{\bar{Q}})^2\,.
\end{equation}
Finally, note that if the charge radius is known, then Eqs.~\eqref{eq:crquark} to~\eqref{eq:crgeneral} can be employed to fix the model parameters.

Figure~\ref{fig:EFFs} shows in the left panel the electromagnetic form factors of the lowest-lying pseudo-scalar charmed mesons, and analogous results for the bottom mesons in the right one. The EFF corresponds to the charged state whose quark-antiquark content match the $q\bar{Q}$ combination where $q$ is the light quark and $\bar{Q}$ the heavy quark, \emph{i.e.} $D^-=d\bar c$, $D_s^-=s\bar c$, $B^{+}=u\bar b$, $B_s=s\bar b$ and $B_c^+=c\bar b$. As a general feature, the meson's electromagnetic form factor falls off with respect the transferred momentum more smoothly when the mass difference of its valence quarks is smaller. In addition, meson's EFFs in the same heavy quark sector seem to decrease asymptotically at the same rate. Note that similar features can be deduced from the results reported in Refs.~\cite{Moita:2021xcd, Das:2016rio, Hwang:2001th, Hernandez-Pinto:2023yin}, we have chosen not to draw them together with ours because the hadronic scale in such models is not specified and it has no sense to perform a detailed comparison at different scales.

Table~\ref{tab:EMradii} collects data from the mentioned references and lattice-regularized QCD calculations on Charge radii of lowest-lying pseudo-scalar heavy-light mesons. We also compare with our findings and, as one can see, our results are in reasonable agreement for the $D_s$, $B_s$ and $B_c$ mesons. However, the charge radii predicted for the $D$ and $B$ mesons are quite large when comparing with other calculations. This should be due to the big difference between the masses of the dressed valence quarks which is not cured by our algebraic model.

\begin{table}[!t]
\caption{\label{tab:EMradii} Charge radii, in fm, of lowest-lying pseudo-scalar heavy-light mesons. Despite the fact that other theoretical approaches do not provide their characteristic low-energy (hadron) scale and therefore a comparison must be made carefully, we have collected the results of the Refs.~\cite{Moita:2021xcd, Das:2016rio, Hwang:2001th, Hernandez-Pinto:2023yin, Li:2017eic, Can:2012tx}.}
\begin{ruledtabular}
\begin{tabular}{lrrrrrr}
& & $D^{-}$ & $D_{s}^{-}$ & $B^{+}$ & $B_s^{0}$ & $B_c^{+}$ \\
\hline
$|r_{0^{-}}|$ & & 0.680 & 0.372 & 0.926 & 0.345 & 0.217 \\
Covariant CQM & \cite{Moita:2021xcd} & 0.505 & 0.377 & - & - & - \\
PM & \cite{Das:2016rio} & -  & 0.460 & 0.730 & 0.460 & - \\
LFQM & \cite{Hwang:2001th} & 0.429 & 0.352 & 0.615 & 0.345 & 0.208 \\
CI & \cite{Hernandez-Pinto:2023yin} & - & 0.260 & 0.340 & 0.240 & 0.170 \\
Lattice & \cite{Li:2017eic} & 0.450(24) & 0.465(57) & - & - & - \\
Lattice & \cite{Can:2012tx} & 0.390(33) & - & - & - & - \\
\end{tabular}
\end{ruledtabular}
\end{table}


\subsection{Impact parameter space GPD}

The IPS-GPD distribution is interpreted as the probability density of finding a parton with momentum fraction $x$ at a transverse distance $b_{\perp}$ from the meson's center of transverse momentum. It can be obtained straightforwardly by carrying out the Fourier transform of the zero-skewness GPD:
\begin{eqnarray}
u_{0^-}(x,b_{\perp}^2)= \int_0^{\infty}\frac{d\Delta}{2\pi }\Delta J_0 (b_{\perp} \Delta) H_{0^-}^q(x,0,t) \,,
\end{eqnarray} 
where $J_0(z)$ is the zeroth Bessel function of the first kind. 

Exploiting the following representation of the GPD in the DGLAP region, 
\begin{equation}
H_{0^-}^q(x,0,t) = q_{0^-}(x) \exp[t \hat{f}_{0^-}^q(x)] \,,
\end{equation}
we can obtain an analytic expression for the IPS-GPD distribution,
\begin{eqnarray}
\label{eq:IPSux}
u_{0^-}^q(x,b_{\perp}^2)= \frac{q_{0^-}(x)}{4 \pi \hat{f}_{0^-}^{q}(x)} \text{exp}\left[-\frac{b_\perp^2}{4\hat{f}_{0^-}^{q}(x)}\right]\;.
\end{eqnarray}
which contains an explicit dependence on the PDF and then reveals a clear interrelation between parton's momentum and spatial distributions within a meson.

Furthermore, considering the mean-squared transverse extent (MSTE),
\begin{eqnarray}
\textless  b_\perp^2(x) \textgreater_{0^-}^q &=& \frac{1}{r_{0^-}}\int_0^\infty db_\perp\, b_\perp^2\, \textbf{b}_{0^-}^q(x,b_\perp) \,, \\
\textbf{b}_{0^-}^q(x,b_\perp) &:=& 2\pi r_{0^-} b_\perp u_{0^-}^q(x,b_\perp) \,. \label{eq:bqblack}
\end{eqnarray}
the IPS-GPD, defined in Eq.~\eqref{eq:IPSux}, yields the plain relation:
\begin{equation}
\textless  b_\perp^2 \textgreater_{0^-}^q = 4\int_0^1 dx\,\hat{f}_{0^-}^{q}(x)q_{0^-}(x) \,.
\end{equation}
Integrating over $x$, and comparing with Eq.~\eqref{eq:crgeneral}, one is left with a compact expression for the expectation value:
\begin{equation}
\textless b_\perp^2 \textgreater_{0^-}^q =  \frac{2}{3}r_{0^-}^2  \left[ \frac{(r_{0^-}^q)^2}{e_q(r_{0^-}^q)^2+e_{\bar{Q}}(r_{0^-}^{\bar{Q}})^2} \right] \,, \label{eq:aveMSTE}
\end{equation}
\emph{i.e.} the expectation value of the MSTE of the valence quark is directly correlated with the meson charge radius.

\begin{figure}[!t]
\centerline{\includegraphics[clip, trim={1.0cm 0.0cm 0.0cm 0.0cm}, width=0.40\textwidth]{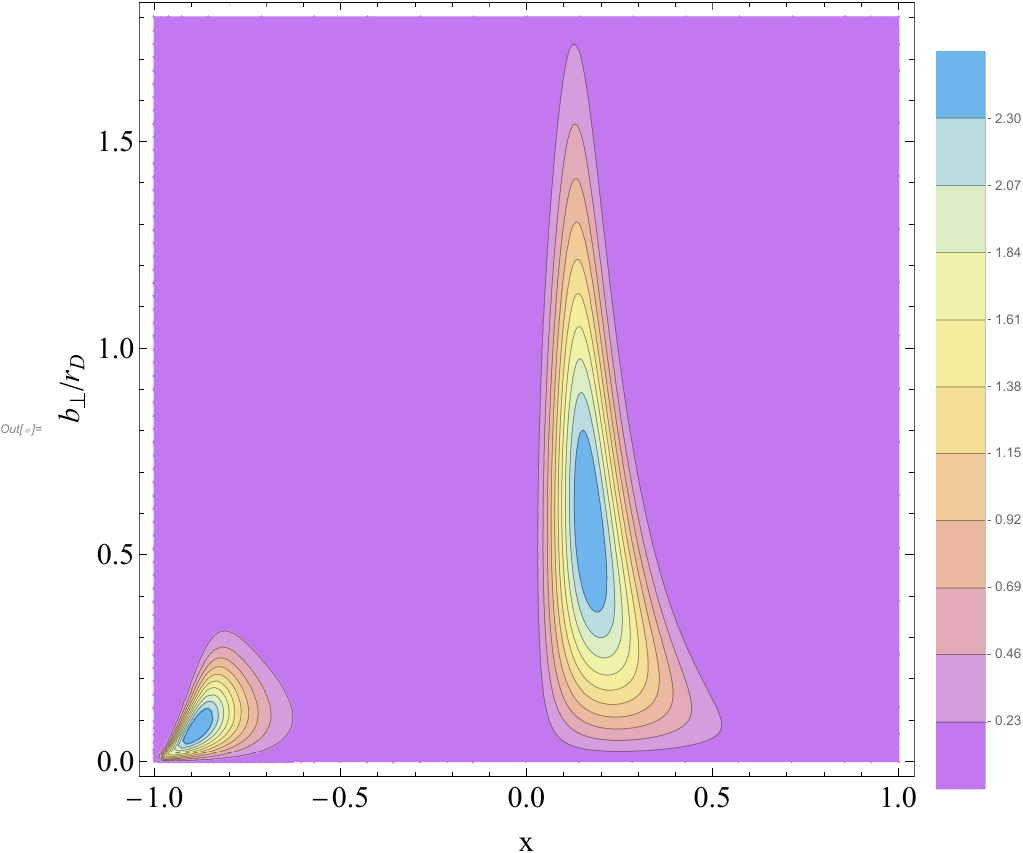}}
\vspace*{0.40cm}
\centerline{\includegraphics[clip, trim={1.0cm 0.0cm 0.0cm 0.0cm}, width=0.40\textwidth]{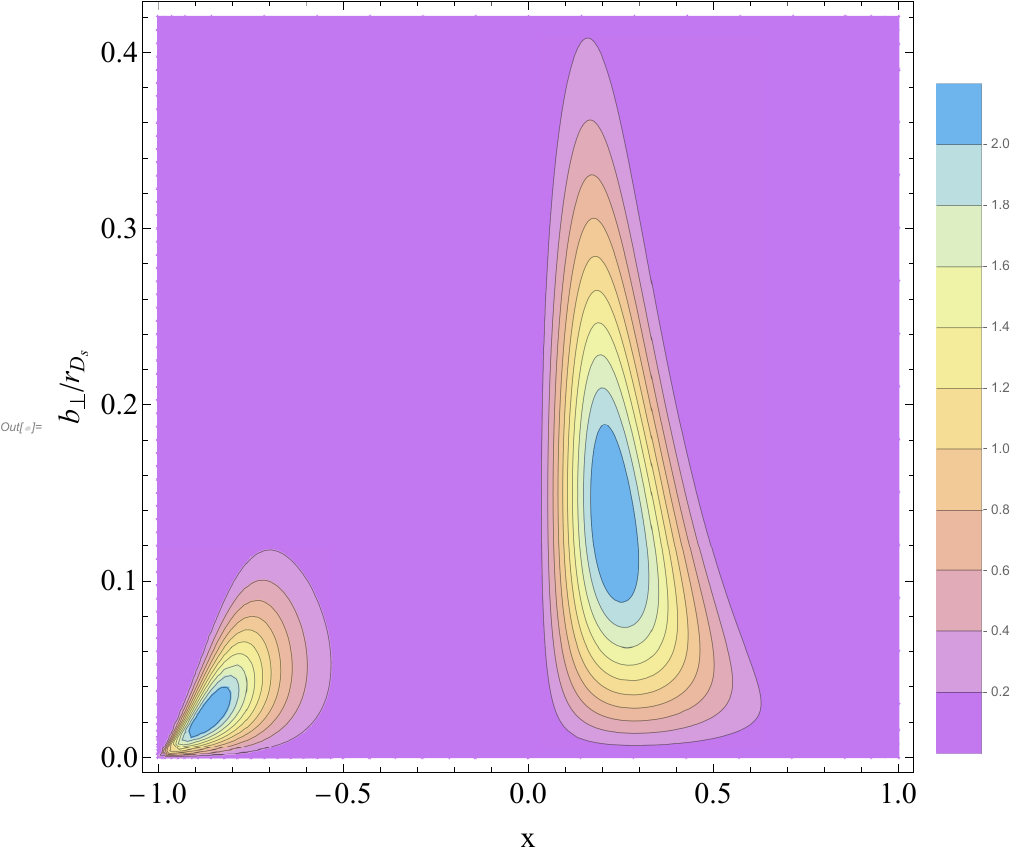}}
\caption{\label{fig:IPS-GPD_Dmesons} Impact parameter space GPDs. For illustrative purposes, we have considered the convenient representation of Eq.~\eqref{eq:bqblack}, where the quark lies in the $x>0$ domain, while the antiquark in $x<0$. \emph{Upper panel:} $D$-meson results. \emph{Lower panel:} analogous results for the $D_s$ meson. The conspicuous asymmetry in both cases is due to the difference in mass of the dressed valence quarks: the heavy quark plays a larger role in determining the center of transverse momentum.}
\end{figure}

Figure~\ref{fig:IPS-GPD_Dmesons} shows the IPS-GPD for the $D$-meson in the upper panel and for $D_s$-meson in the lower one. For illustrative purposes, we have considered the convenient representation of Eq.~\eqref{eq:bqblack}, where the quark lies in the $x>0$ domain, while the antiquark in $x<0$. One can conclude that the heavy antiquark is almost fixed at the center of transverse momentum whereas the highest probability of finding the light quark in the transverse plane is at a distance $0.6\times r_D$ and $0.14\times r_{D_s}$ for the $D$ and $D_s$ mesons, respectively. It is also seen that as the constituent quark mass becomes larger the quark plays an increasingly major role in determining the center of transverse momentum; furthermore, the distributions become wider in $x$ but less extended in $b_\perp$, and their maximum becomes smaller.

\begin{figure}[!t]
\centerline{\includegraphics[clip, trim={1.0cm 0.0cm 0.0cm 0.0cm}, width=0.38\textwidth]{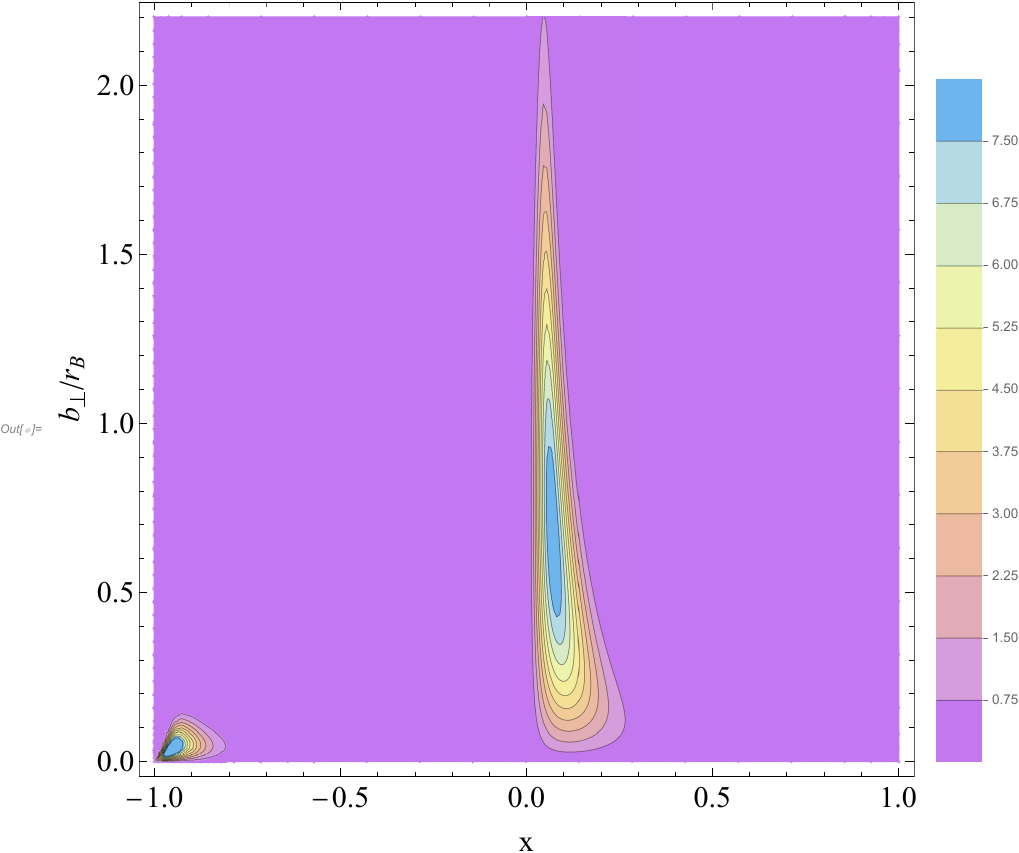}}
\vspace*{0.60cm}
\centerline{\includegraphics[clip, trim={1.0cm 0.0cm 0.0cm 0.0cm}, width=0.38\textwidth]{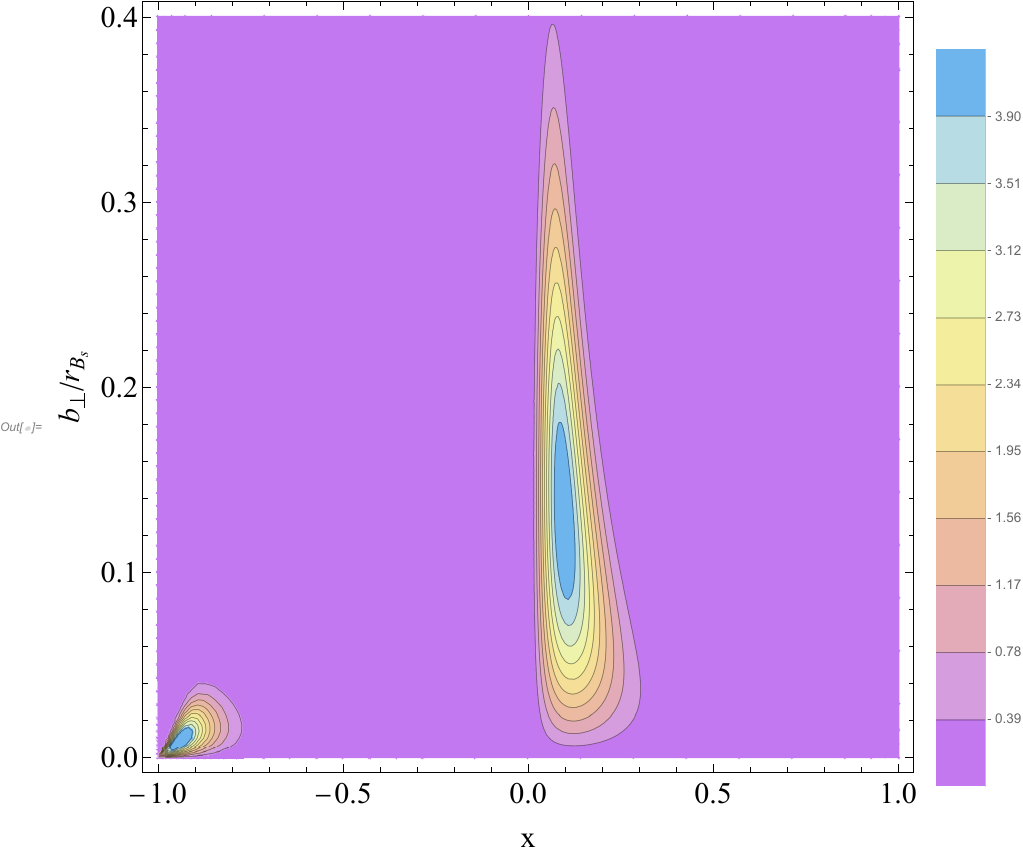}}
\vspace*{0.60cm}
\centerline{\includegraphics[clip, trim={1.0cm 0.0cm 0.0cm 0.0cm}, width=0.38\textwidth]{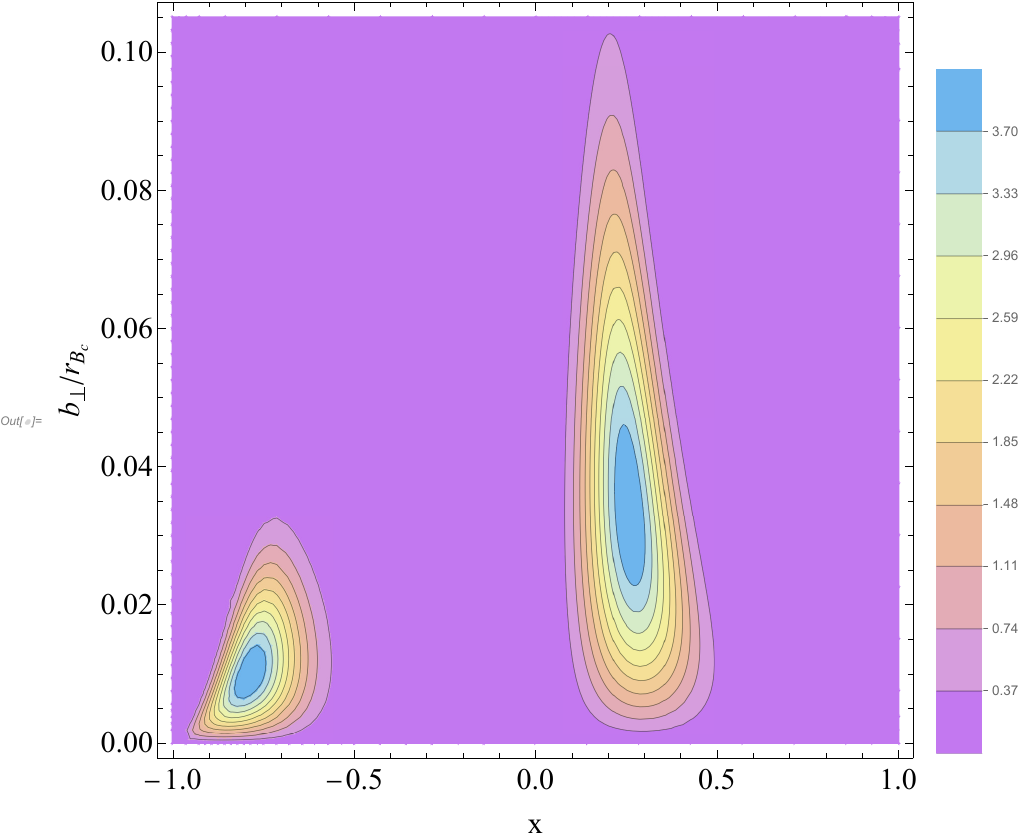}}
\caption{\label{fig:IPS-GPD_Bmesons} Impact parameter space GPDs. For illustrative purposes, we have considered the convenient representation of Eq.~\eqref{eq:bqblack}, where the quark lies in the $x>0$ domain, while the antiquark in $x<0$. \emph{Upper panel:} $B$-meson results.  \emph{Middel panel:} $B_s$ results \emph{Lower panel:} $B_c$ results. The conspicuous asymmetry in all cases is due to the difference in mass of the dressed valence quarks: the heavy quark plays a larger role in determining the center of transverse momentum.}
\end{figure}

Figure~\ref{fig:IPS-GPD_Bmesons} shows analogous IPS-GPD results for the $B$-, $B_s$ and $B_c$ mesons in the upper, middle and lower panels, respectively. Similar features as the ones highlighted for the charmed mesons can also be guess for the bottom mesons. Since the quark lies in the $x>0$ domain, while the antiquark in $x<0$, one can observe that the heavy antiquark determines the center of transverse momentum, this time even strongly due to the mass difference between the light quark and the heavy antiquark that constitute the bottom meson. Moreover, the light quark is more frequently located in the transverse plane at a distance $0.65\times r_B$, $0.13\times r_{B_s}$ and $0.035\times r_{B_c}$ for the $B$, $B_s$ and $B_c$ mesons, respectively. Note here the fact that the $c$-quark and $b$-antiquark are quite close in the transverse plane for the $B_c$-meson. Besides, as in the charmed sector, the distributions become wider in $x$ but less extended in $b_\perp$ when the constituent quark mass becomes larger; this feature also affects the center of transverse momentum and the maximum values reached by the distributions.


\section{Summary}
\label{sec:summary}

Following a recently proposed algebraic model which adequately describes the internal structure of the lowest-lying hidden-flavor pseudo-scalar and vector mesons, with either light or heavy quark content, we have applied the same idea to the heavy-light pseudo-scalar mesons. The advantage is always the same, the knowledge of the leading-twist LFWF from its connection to the corresponding valence-quark PDA which is usually more amenable to compute within sophisticated continuum approaches to the valence-quark and -antiquark bound-state problem in QCD.

The algebraic model consists on constructing simple and evidence-based \emph{ans\"atze} of the meson's BSA and quark's propagator in such a way that the BSWF can then be readily computed. Its subsequent projection onto the light front yields the light front wave function (LFWF) whose integration over the transverse momentum squared permits an algebraic connection with the PDA. 

We have exploited our current accurate knowledge of the PDAs for lowest-lying heavy-light pseudo-scalar mesons to determine the corresponding LFWFs whose overlap representation delivers the GPDs in the DGLAP region. With the GPDs at hand, performing different limits/projections, the related PDFs, EFFs and IPS-GPDs of the mesons under study have been computed. Experimental results and earlier theoretical predictions are scarce and incomplete; therefore, we have provided comparison when possible but this work should motivate future experiments and theoretical developments.


\begin{acknowledgments}
B. Almeida Zamora acknowledges CONACyT (No. CVU 935777) for PhD fellowship.
J.J. Cobos-Martínez acknowledges financial support from the University of Sonora under grant USO315007861.
%
A. Bashir acknowledges Coordinaci\'on de la Investigaci\'on Cient\'ifica of the Universidad Michoacana de San Nicol\'as de Hidalgo Grant No. 4.10., the US Department of Energy (DOE) under the Contract No. DE-AC05-6OR23177 and the Fulbright-Garc\'ia Robles Scholarship.
This work has also been partially funded by Ministerio Espa\~nol de Ciencia e Innovaci\'on under grant No. PID2019-107844GB-C22; Junta de Andaluc\'ia under contract Nos. Operativo FEDER Andaluc\'ia 2014-2020 UHU-1264517, P18-FR-5057 and also PAIDI FQM-370.
\end{acknowledgments}


\bibliography{PrintLFWFsPseudoscalarHeavyLight}

\end{document}